# GCTAg: scalable mixed-model analysis for biobank-scale agricultural cohorts

Alexander S. Leonard[1,*], Qiongyu He[1], Natasha Watson[1], Naveen Kumar Kadri[1], Hubert Pausch[1]

[1] Animal Genomics, ETH Zurich, Universitaetstrasse 2, 8092 Zurich, Switzerland
* alleonard@ethz.ch

**Abstract**
Genome-wide association studies identify genomic variants associated with traits. Mixed linear model association (MLMA) methods using a whole-genome relationship matrix, such as those implemented in GCTA, are powerful but computationally expensive. Here, we remove key memory and CPU bottlenecks in GCTA, reducing REML memory usage by nearly 75% and substantially accelerating MLMA by orders of magnitude in biobank-scale cohorts while preserving exactness. We further exploit relatedness in the mapping cohort through a reduced-rank Woodbury matrix approach, delivering further orders of magnitude performance gains with controlled genomic inflation. Native on-the-fly dominance recoding also eliminates slow I/O-operations on intermediate files, enabling efficient additive and dominance MLMA analyses in large agricultural cohorts.

**Introduction**
Genome-wide association studies (GWAS) are a key discovery step in identifying variants which are associated with specific phenotypes or traits. Mixed linear models emerged as a method of choice for GWAS as they minimize spurious associations by accounting for confounding factors like population stratification and relatedness within the mapping cohort [1,2]. GCTA (Yang et al. 2011, 2014) remains one of the most popular software tools for GWAS and mixed linear model associations (MLMA). Successful MLMA requires careful handling of genomic relatedness matrices (GRM) and estimating variance components from the null mixed model before testing each variant for association with traits and diseases. Due to the heavy use of matrix operations, efficient use of linear algebra libraries can vastly accelerate these calculations [4]. However, MLMA remains computationally demanding and tends to scale poorly to large mapping cohorts comprising several hundred thousand individuals genotyped at millions of markers. The development of association testing methods that appropriately account for sample relatedness while maintaining scalability to increasingly large mapping cohorts has become an active area of research in the past decade.

Several algorithmic advancements in MLMA largely depend on exploiting "sparse" GRMs, where individuals are minimally related and off-diagonal GRM entries can be neglected. Software tools such as BOLT-LMM [5], GCTA-fastGWA [6], and REGENIE [7] can scale to much larger cohorts than GCTA-MLMA. However, agricultural populations tend to be highly related due to small effective population size, strong artificial selection and the widespread use of a small number of elite breeding individuals. As a result, mapping cohorts in these populations typically have "dense" GRMs and can invalidate the sparse GRM assumption of such tools, inflating the test statistics in many agricultural GWAS [8,9]. With an increasing number of individuals being genotyped on dense arrays [10] due to national genotyping initiatives [11–13], scalable methods that account for the typical structure of agricultural populations are needed to fully leverage the power of biobank-scale cohorts for complex trait analysis.

One common approach for handling large GWAS cohorts is to split samples into multiple smaller subsets for MLMA followed by a meta-analysis, as recently applied [14–16]. Alternatively, modelling all variants with a fixed variance term [12] or stochastic REML methods (i.e. as implemented in SLEMM [17]), can scale to much larger cohorts. However, these approaches can be less robust when modelling large-effect SNPs or may implicitly fail to converge to stable numerical solutions [18]. Other approaches build on the Algorithm for Proven and Young (APY), a reduced-representation appropriate to dense GRMs using a smaller set of "proven" samples to represent the larger set of "young" samples [19]. Similar reduced-rank approaches help ease the matrix inversion bottlenecks in variance components estimation [20] or Genomic Best Linear Unbiased Prediction (GBLUP) [21,22].

GWAS typically assume additive effects of quantitative trait loci (QTL), where a heterozygous genotype contributes half the effect of a homozygous alternate genotype. While biobank-scale studies in human populations indicate that genetic variance for complex traits is predominantly additive [16], there is increasing evidence from large-scale studies in livestock that dominance (i.e. a subset of non-additive) QTL can remain undetected when the statistical models assume additive inheritance [23–27]. Despite dominance "genotypes" simply being a numeric recoding [28], testing these alternative modes currently requires generating an additional set of genotypes in formats such as the cumbersome MaCH dosage format [29]. The MaCH dosage format has previously been used to represent genotype probabilities and account for uncertainties introduced by genotype imputation [30,31].

Here, we reimplemented several core components of GCTA to increase GWAS throughput and improve memory management, as well as several approximations in the PCA/REML steps. We also implemented several new features including on-the-fly dominance recoding to further streamline and expand the functionality. Finally, we demonstrate that these modifications now allow using "GCTAg" (GCTA with improvements targeted at agricultural species compatible with the GPL-3 license) for association testing and related investigations in cohort designs that previously required subsetting and meta-analysis strategies.

## Results

To benchmark GCTA and our reimplementation GCTAg, we first simulated a large genotype-phenotype dataset consisting of $N$=152,522 individuals genotyped at $m$=600,000 markers. We used AlphaSimR to seed a founding population followed by generations of selection and mating to mimic cattle populations. We simulated a moderately heritable phenotype with genetic variation arising from additive effects at several distributed QTL. Using subsets of this dataset (from $N$=9,375 to $N$=150,000) we benchmarked typical GCTA GWAS for both runtime and peak memory. We benchmarked these analyses on different compute architectures, using an Apple M5 laptop with 10 cores and 24 GB of RAM ("Laptop") and Linux AMD EPYC 9654 compute nodes with 96 cores and 384-768 GB of RAM ("Server").

The full-GRM mixed linear model GWAS implemented in GCTA is comprised of four main steps: (i) building the genomic relationship matrix (GRM) from standardized genotypes; (ii) computing its principal components by eigendecomposition; (iii) estimating the genetic and environmental variance components by REML; and (iv) testing each variant against the covariate-adjusted null model. We decoupled the REML and MLMA steps in GCTA, implementing a lossless save/load feature for the REML result. This allowed cleaner benchmarking of REML (typically memory-bound) and MLMA (typically compute-bound). We profiled CPU utilisation and memory allocation using this minimally modified version of GCTA on Laptop and Server on a reduced $N$=9,375 $m$=150k subset of the simulated data.

We observed that the GRM and PCA steps were largely dispatched immediately to linear algebra operations (**Supplementary Figure 1**). On the other hand, the REML and MLMA steps spent appreciable time in data input/output (I/O) (**Supplementary Figure 2**). The REML step also suffered from excessive memory churn from reallocating the projection and phenotypic covariance matrices (**Supplementary Figure 3**) while the MLMA step was bottlenecked in SGEMV (**Supplementary Figure 2**). We first sought to address these bottlenecks in GCTAg (**Supplementary Figure 1**), then expand on existing functionality, and finally add several approximations to scale to significantly larger cohorts.

*Exact implementation improvements*

Although the GRM step was almost entirely spent within the highly efficient DGEMM BLAS-3 operation, it only batched (a hardcoded) 128 variants at a time. Depending on the CPU and core count, increasing the number of variants considered during each GRM accumulation step helps shift the kernel from memory-bound to compute-bound, achieving a noticeable increase in throughput (**Figure 1A**, **Supplementary Figure 4**). This was especially important at high core counts, as commonly found on HPCs, gaining 30+% performance for a minor increase in memory. However, since the GRM is symmetric, we can explicitly only solve one triangle of the matrix and avoid allocating memory for the other triangle, leading to an effective ~35% reduction in peak memory (**Supplementary Note 1**). Almost all GWAS steps can exploit this upper/lower triangular form, leading to general memory savings.

The default SelfAdjointEigenSolver used in GCTA for the eigenvalue decomposition dispatches to the "QR iteration" algorithm (DSYEV in LAPACK), as opposed the newer and typically faster "Divide and Conquer" (D&C) algorithm (DSYEVD). Both algorithms calculate the entire eigendecomposition, and so when instructing GCTA to calculate only the top $k$ eigenvalues (such that $k \ll N$ is not the full rank of the GRM), we instead used the "Multiple Relatively Robust Representations" (MRRR) algorithm (DSYEVR) to calculate only the top $k$ eigenvalues more cheaply (**Figure 1B**, **Supplementary Figure 5**). In particular, the DYSEV algorithm appears to have a non-reference implementation in the Intel Math Kernel Library (MKL) and is much less performant in other LAPACK libraries (**Supplementary Note 2**), whereas we can largely close the performance gap across using the D&C or MRRR algorithms. This is especially important for ARM-based Apple machines which cannot use MKL.

During the REML step, we were also able to fully exploit the symmetric matrix $V$, only filling the lower triangle required by the Cholesky factorization, reducing memory bandwidth. We also reformulated the covariate correction of the projection matrix as a symmetric rank-c update, which avoids explicitly forming the covariate-adjusted projection matrix. By aggressively freeing, solving in-place, or swapping large matrix buffers, we were also able to markedly reduce peak memory during REML by almost 75% and runtime by 33% (**Figure 1C**).

The largest computational bottleneck was the variant scan in the MLMA step, where covariate-adjusted test statistics were computed one variant at a time. Each per-variant operation was memory-bound (BLAS-2) and too small to parallelize efficiently. We restructured this step to process variants in blocks, using globally transformed phenotypes and genotypes, so that the covariate adjustment is applied once rather than repeated for every variant. By replacing many small matrix-vector operations with a few large matrix-matrix operations (BLAS-3) and only computing necessary terms (e.g. on the diagonal), we decreased runtime by almost 99% (**Figure 1D**) while preserving mathematically exact test statistics. Given BLAS-3 has better multicore scaling than BLAS-2, this gap would likely continue to grow at higher core counts.

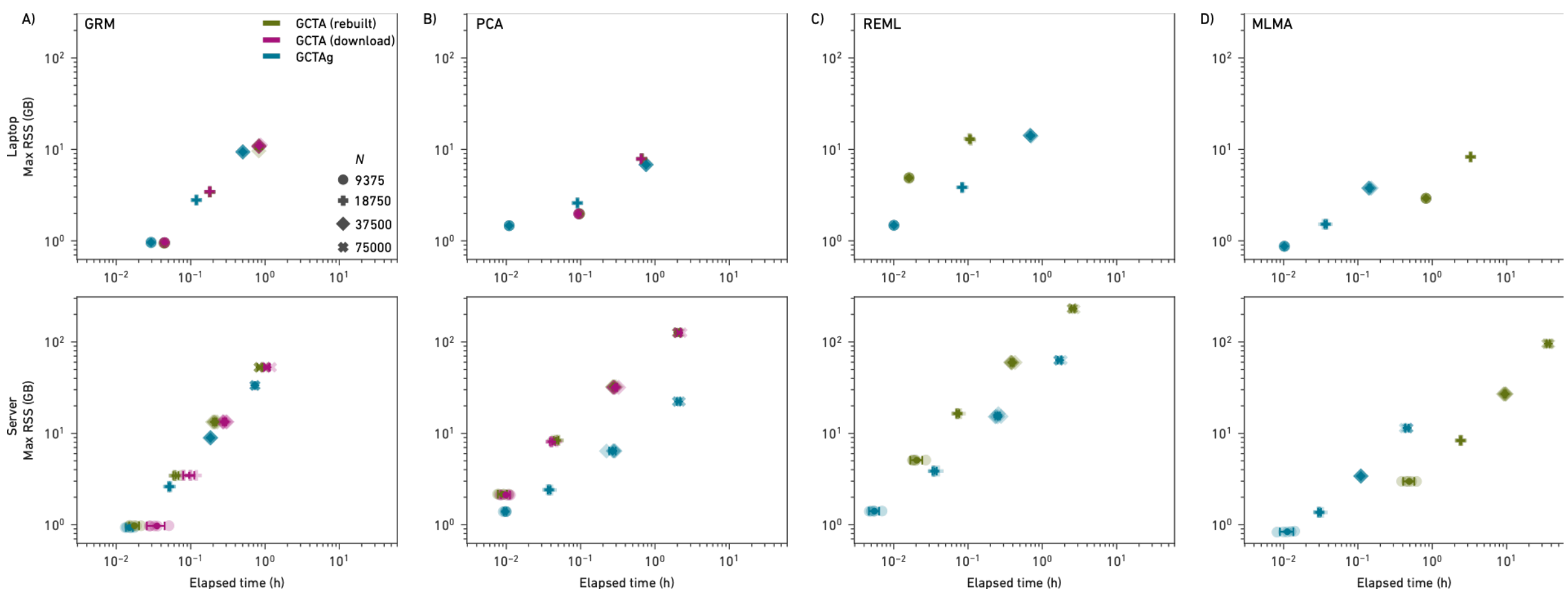

*Figure 1. GCTA versus GCTAg compute resources used during the (A) GRM, (B) PCA, (C) REML, and (D) MLMA steps for four datasets of increasing sample size tested on Laptop using 10 cores (top) and Server using 24 cores (bottom). Each run was replicated five times (faint markers). GCTA (rebuilt) refers to our minimally modified GCTA version, while GCTA (download) is the unmodified publicly available binary. The REML and MLMA steps are not decoupledq in the downloaded version and were not run. The Laptop had insufficient memory to run GCTA for the* N=37,500 *dataset for the PCA, REML, and MLMA steps.*

Although data I/O is typically a small contributor to overall runtime for large datasets, we also identified bottlenecks where reading/writing data was typically done byte-by-byte. This was vastly accelerated using memory-mapped files, contiguous memory allocation, and buffered reads/writes (**Supplementary Figures 2 and 3**). However, since the test statistics are calculated blockwise, we can also naturally stream blocks of genotypes with no loss of BLAS efficiency. By avoiding loading all data upfront, and only unpacking a block of variants, we can reduce peak memory usage substantially (from $\mathcal{O}(m)$ to $\mathcal{O}(1)$ memory). Just unpacking the genotype data at $N$=150k, $m$=600k requires 90 GB of memory, compared to processing batches of 1 GB. For per-chromosome MLMA (using $m$=600k variants), we reduced peak memory by almost 90% (**Figure 1D**).

We refactored GRM merging, replacing the naïve implementation of sequentially adding each GRM element with four dense matrices in memory (GRM itself and SNP count for the per-iteration and accumulated copies). Since the lower-triangle binary file format of the GRMs is effectively one-dimensional, we can instead accumulate and then write chunks across all files, without ever holding the dense matrix in memory. This approach uses only ~0.5 GB (critically, independent of GRM size) and 15 minutes at *N*=150k with 4 cores (cf. 615 GB and 10 hours for GCTA) (**Supplementary Figure 6**). These changes were especially crucial to prevent GRM-merging from becoming the top memory bottleneck after the other optimisations.

*Exact robustness extensions*

We also formalised the REML convergence criterion and improved the variance components initialisation. We replaced the “backward-looking” criterion, stopping when the change in log-likelihood between iterations fell below a fixed threshold. Now, convergence is based on a “forward-looking” criterion, stopping when the remaining optimisation signal, estimated from the Average Information (AI) matrix as a Newton decrement, falls below a fixed tolerance. We also refined the step-size damping applied to the variance components updates, taking full steps when the model is trustworthy and damping otherwise. Across three simulated traits, this resulted in converging in around half as many rounds (i.e. 50% reduction in REML runtime) with negligible difference in heritability estimates (**Supplementary Table 1**, **Supplementary Figure 7**).

We also reimplemented the initial seeding of the variance components, using a Haseman–Elston (HE) estimator [32] to make a data-driven prediction as opposed to assuming an equal partitioning of variance. Because the HE starting point is numerically stable for any reasonably conditioned GRM, REML can directly proceed with the quadratically converging AI algorithm and skip the initial round of linearly converging Expectation–Maximization algorithm. The HE seeding raises the initialization cost from negligible to $O(n^2)$, but since each EM- or AI-REML iteration is $O(n^3)$, eliminating even one iteration typically yields a net saving. For Trait A, additive with low heritability, the improved initialisation reduced REML rounds by almost 60% (**Supplementary Table 1**, **Supplementary Fig. 7**).

GCTA natively supports partitioning the GRM building (using --make-grm-part), which becomes important at large sample sizes (since GRM memory scales as $\mathcal{O}(n^2)$). However, for situations that are memory-bound rather than compute-bound (i.e. running on a single machine or a small cluster), we implemented an internal tiled GRM building approach. By specifying a memory budget, we can dynamically calculate as many GRM rows as fit in the budget. Calculating the entire GRM this way was comparable in time to the partitioned approach (but from a single invocation and no subsequent file merging) and was negligibly slower than the full-build approach when specifying a budget around 50% of the memory used by the full-GRM (**Supplementary Figure 8**). Reducing the budget further incurs more time-consuming I/O passes but enabled calculating an *N*=75k GRM using 90% less memory while only being 26% slower.

Given the frequent p-value underflow for large cohorts and/or strong associations, as the p-value approaches $2^{-1022} \approx 2.2 \times 10^{-308}$, we also added the option to report the logarithm of the significance, using an asymptotic expansion of the Regularized Upper Incomplete Gamma function: $\ln Q(a, x) \approx -x + (a-1)\ln x - \ln\Gamma(a)$ for $x \gg a$. This allows us to accurately report more significant associations, as well as naturally forming the more commonly reported $-\log_{10}(p)$ significance. This allows exact reporting of significances as GWAS scale to larger cohorts that potentially produce vanishingly small p-values.

We also generalized the code to use any BLAS (Basic Linear Algebra Subprograms) and LAPACK (Linear Algebra PACKage) libraries using standard function notation where necessary. The choice of BLAS/LAPACK backend can have noticeable impact, e.g. using the Apple-provided Accelerate SDK was 45% faster than the OpenBLAS library and used 30% less memory under otherwise identical conditions on Laptop (**Supplementary Figure 9**). On Server, using AMD Optimizing CPU Libraries (AOCL) was under 2% faster than OpenBLAS (the next fastest library tested) with similar memory usage with 6 cores. However, AOCL is tuned specifically for the AMD nodes used by Server, and was significantly better at parallelizing across more cores, almost 70% faster than OpenBLAS and 92% faster than MKL with 192 cores (**Supplementary Figure 9**). However, the MKL version is substantially faster for most LAPACK routines (**Supplementary Note 2**), and so we linked AOCL and MKL for BLAS and LAPACK respectively (with AOCL taking precedence over MKL for BLAS routines). CPU architecture can also have a significant impact on performance, with Zen 4-based server nodes almost twice as fast on average as a Zen 2-based node under similar conditions (**Supplementary Figure 10**). This improvement is largely driven by wider SIMD (single instruction, multiple data) support on Zen 4 CPUs.

Loading the GRM into memory, even with the ~50% savings with the triangular form, still grows quadratically with sample size. Several REML quantities can be computed "out-of-core", where the result is accumulated over blocks of the GRM, allowing us to only load a fraction of the GRM into memory at any given time. This reduces peak memory requirements (**Supplementary Figure 11**) but can increase I/O pressure substantially (both through repeated reading of the GRM across REML iterations as well as less efficient smaller reads). Although reduced, the memory scaling is still quadratic due to other terms which cannot be easily distributed.

*Approximate extensions enable scaling to larger cohorts*

A primary bottleneck in REML iterations is calculating the trace of the projection matrix product with each variance component's matrix. We replaced this exact calculation with a stochastic Hutch++ trace estimator [33], which never requires explicitly forming the projection matrix and reduces the cost from $O(n^3)$ to $O(sn^2)$ for $s$ probe vectors. This reduced REML runtime and memory by 67% and 16% respectively compared to the exact mode (**Figure 2A**), had no effect on MLMA runtime (**Figure 2B**), and was within 0.1% relative genomic inflation to the exact mode (**Figure 2C**). More probes reduce the standard error at the cost of runtime, but too few can impede convergence and cancel out the gains (**Supplementary Figure 12**). Although the Hutch++ estimator is itself unbiased, we observed small deviations above and below the exact test statistics because REML applies non-linear operations to the estimated traces.

We also implement two additional PCA methods, Lanczos and randomized singular value decomposition (rSVD). Both methods are highly efficient at extracting a modest number of top *k* eigenvalues, and the former is mathematically exact under exact arithmetic while the latter is approximate but highly accurate. They can be orders of magnitude faster than even the MRRR algorithm from LAPACK introduced earlier (**Supplementary Figure 5**). For the top four PCs at *N*=75k, as commonly used as population structure covariates, rSVD reduced the runtime by 99% compared to MRRR, or a 95% reduction for the top 10% of PCs. The rSVD method can also efficiently leverage the out-of-core GRM operations. For rSVD, we implemented two ways of constructing the leading eigensystem: power iterations and a Nyström approximation, with the latter prioritising runtime at the cost of accuracy.

Many recent tools, such as BOLT-LMM and REGENIE, achieve fast association testing through approximations that assume a sparse GRM, in which individuals have negligible relatedness. Agricultural GWAS cohorts often contain highly related individuals (**Supplementary Table 2**), which violates this assumption and can inflate test statistics [8]. However, rather than relatedness being a hindrance, we instead exploited a low-rank approximation of the GRM. Using its leading $k$ eigenpairs and applying the Woodbury matrix identity, the covariance matrix inverse required at each REML iteration involves only a $k \times k$ matrix rather than an $N \times N$ one. The GRM information concentrates into fewer leading components as relatedness increases, leading to massive efficiency gains when $k \ll N$. For $N$=150k and $k$=10k, we reduced REML runtime and memory by 95% and 53% respectively compared to the exact mode. We reused this reduced basis to compute the variant-scan test statistics without ever forming the full inverse, also reducing MLMA runtime and memory by 88% and 85% respectively (**Figure 2B**) for a ~3.5% relative genomic inflation to the exact mode.

We implemented three data-driven methods to select $k$, based on Marchenko–Pastur (MP) theory from random matrices, the eigenmass approach (EIG) from [8], or limiting the variance (VAR) in the discarded tail. We reused the rSVD implementation described previously for solving the eigensystem, initially calculating a small number of eigenvalues and doubling as necessary until the selection method satisfies its stopping conditions. Calculating many PCs is computationally expensive, and so datasets with a slowly decaying eigenspectrum may spend longer calculating the rank $k$ than just running REML with the Hutch++ approximation. The methods depend differently on the distribution of eigenvalues, and so may be overly aggressive or conservative for a given dataset (**Supplementary Figure 13**). In particular, the GRM is not a random matrix, and linkage disequilibrium means variants are not independent variables, and so the MP approach is cheap but tends to significantly underestimate the value of $k$.

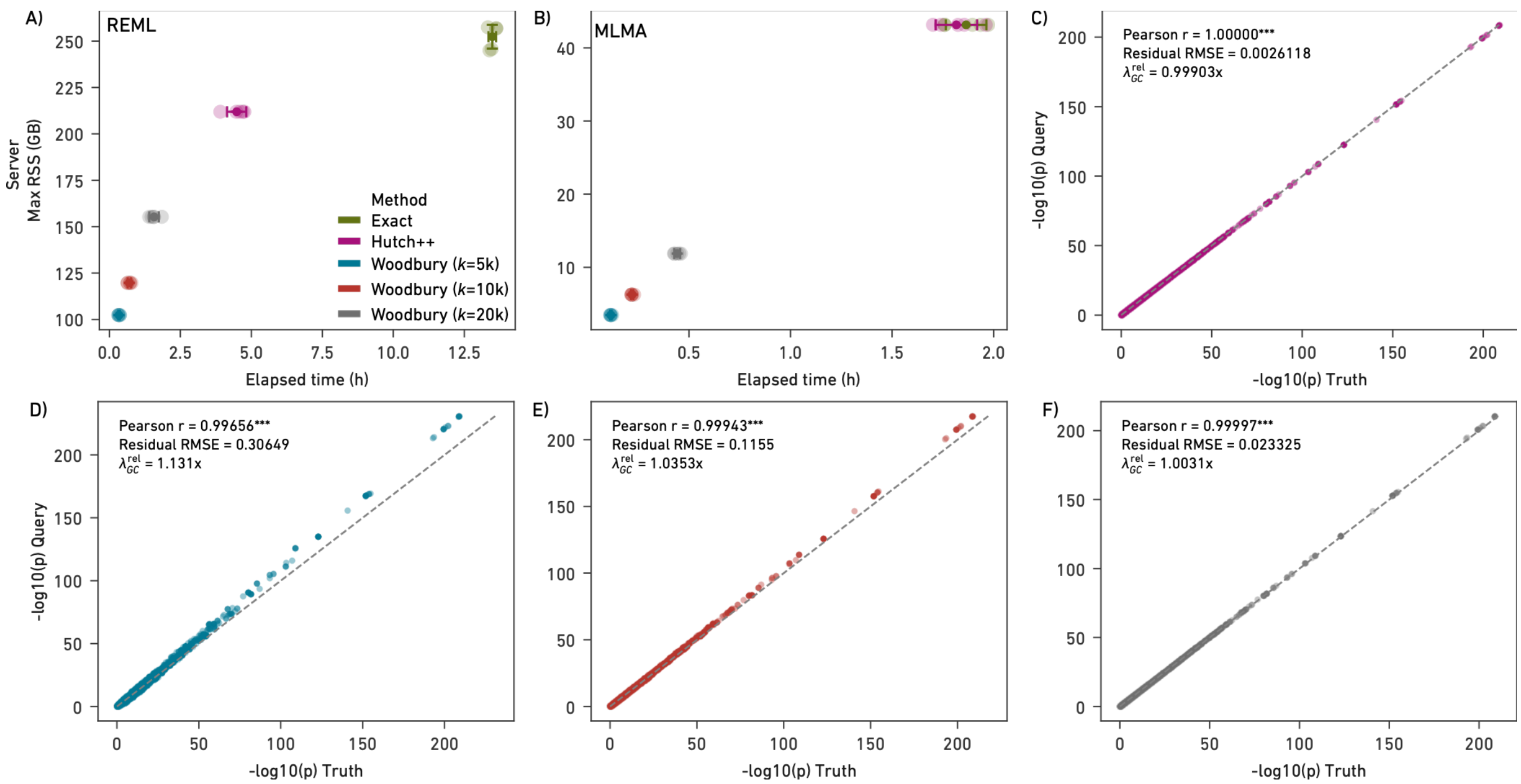

*Figure 2. Comparison of approximations used during (A) REML and (B) MLMA steps using* N*=150k,* m*=600k with 24 cores. Note, methods without "Woodbury" use the full matrix rank, and so are effectively equivalent during the MLMA step (with only minor variation to the REML solution). Error bars indicate the mean and standard deviation across five replicates. (C-F) Variant significances are strongly correlated between the approximate and exact methods and show minimal relative genomic inflation compared to the exact mode. The relative genomic inflation factor $\lambda_{GC}^{rel}$ is the ratio of genomic inflation factors for the exact and approximate methods, where a value greater than 1 indicates the approximate method is more inflated than the exact method.*

While aggressive selection of a small *k* will be faster, the truncation of signal in the eigendecomposition tail biases estimates of heritability (**Supplementary Table 3**) and thus typically inflates test statistics and significances (**Figure 2D**-**F**) with a roughly constant factor. The information from the truncated eigenvalues can be used to partially correct the heritability estimates, but not the REML itself (**Supplementary Note 3**), and increasing *k* is the most reliable approach to improve accuracy.

*Direct support for genotype recoding*

Conventionally, additive QTL are tested using the genotype coding scheme mapping aa, aA, and AA to 0, 1, and 2 respectively. Alternatively, the dominance variance at individual SNPs can be tested by respectively recoding genotypes to 0, 2*p*, and 4*p*-2, where *p* is the allele frequency for that variant [28]. A previous modification of GCTA, used in the recent large-scale GWAS of the 1000 Bull Genomes project [30], enabled using the already supported MaCH dose/info file format in MLMA. Previous studies further repurposed this machinery by setting the dosage to the dominance recoding [24]. However, the MaCH format is "sample-major" whereas PLINK binary BED files (typically the primary data input source for GCTA and other GWAS tools) are "variant-major". Converting between BED and MaCH is computationally expensive in both directions in terms of CPU and memory (**Figures 3A** and **3B**), as well as requiring intermediate disk storage (**Figure 3C**). MaCH is also text-format, and so can be compressed to lower storage footprint but incurring additional compression/decompression time. For the *N*=150k dataset across all 29 chromosomes (*m*=17.4M) variants, this would require an additional 6 TB of storage for the compressed MaCH format. Furthermore, the round-trip encoding of floating-point numbers as fixed-length strings in a text file is not lossless, compared to GCTAg's native binary file decoding.

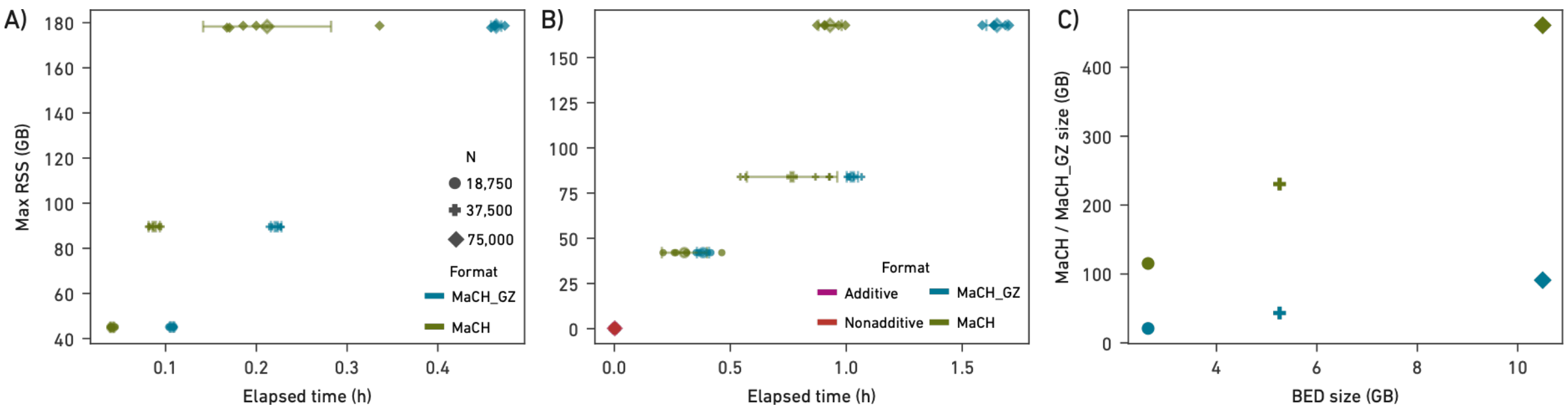

*Figure 3. Genotype recoding using 12 cores. (A) Recoding PLINK binary BED files to plain-text MaCH (MaCH) or compressed MaCH (MaCH_GZ) is a memory-intensive step. (B) Recoding genotypes on-the-fly from the PLINK BED file is substantially faster than loading MaCH or MaCH_GZ data, due to avoiding reshaping of the variant- or sample-major axis and the ability to stream genotypes. Since we can load and process data in parallel when streamed, the step is largely limited by I/O speed (on the order of GB/s). (C) The bit-packed format of PLINK BED genotypes requires a significantly smaller storage footprint compared to text-based MaCH, even when compressed, and does not require intermediate files on disk.*

Furthermore, the MaCH format cannot be easily streamed, as now implemented for BED files, requiring a large memory footprint to load and transpose the numeric genotypes. Even after optimizing a BED-to-MaCH conversion script and the GCTA MaCH loading paths, these two steps are typically the largest bottleneck and can dominate the runtime of the entire MLMA step. By implementing a recoding on-the-fly strategy, we vastly simplified the process of testing the dominance variance at individual SNPs. Crucially, this still allows subsetting variants or samples, which can affect the allele frequency estimates, whereas MaCH files would have to be regenerated on disk for any subset.

**Discussion**

GCTA remains an integral tool for association testing, especially for datasets which do not satisfy the assumptions of fast methods like BOLT or REGENIE. Even tools like SLEMM, targeted towards testing massive livestock cohorts, do not replace full-GRM mixed linear models like that in GCTA [8]. Here, we both improve the existing GCTA implementation and extend with validated approximations to substantially accelerate the GWAS workflow. These changes also better leverage current computing infrastructure (SSDs, SIMD, larger CPU caches, etc), and the ability to split memory- or compute-bound steps for optimized use of HPC or cloud computing resources. Such improvements have a substantial impact on both the feasibility and compute/carbon footprint considerations of large scale GWAS [33]. This mirrors similar developments in other bioinformatic software, for example the ubiquitous short read aligner bwa [34], with e.g, bwa-mem2 and bwa-mem3 [35], bwa-meme [36], etc increasing throughput primarily through engineering/implementation, as opposed to minibwa [37] using a fundamentally different algorithmic design.

The improved complexity and parallelization of the variant-scan in the MLMA step is especially crucial. As "sequence-imputed" variants, typically representing whole genome variant panels with tens of millions of variants, gain popularity over marker panels, on the order of tens or hundreds of thousands of variants, GWAS designs need to scale appropriately. By entirely removing the variant-scan bottleneck and streaming the genotypes, GWAS with GCTAg are now primarily limited (in terms of both runtime and peak memory) in the REML step by the number of samples considered rather than number of variants. At large (*N*=150k, *m*=600k) cohort scales, even with the memory optimizations, REML with GCTAg required around 268 GB of memory and 13.5 hours and MLMA required 45 GB of memory and 1.9 hours with 24 cores. The runtime for matrix inversions at each REML iteration scale cubically with samples (and quadratically for memory, even with out-of-core GRM operations), and so is impracticable to scale much larger while retaining exactness.

To support scaling to even larger cohorts, we implemented two distinct approximations. The Hutch++ approximation scales quadratically with samples for runtime and avoids materialising $P$, with REML using around 220 GB of memory and 4.5 hours for the large cohort (and MLMA unchanged). The approximation is highly accurate, with only minor unbiased stochastic noise in the test significances (relative genomic inflation under 0.1%). The reduced-rank Woodbury basis approximation scales substantially more favourably with sample size, requiring 126 GB of memory and only 42 minutes for *k*=10k (relative genomic inflation of ~3.5%). Notably, the Woodbury basis approximation also impacts the MLMA step, using only 7 GB of memory and 13 minutes. Holding the GRM in memory becomes especially dominant in the Woodbury mode, and so out-of-core GRM operations allows us to reduce REML memory to just 54 GB (-57%) with an acceptable increase in runtime to 65 minutes (+55%) runtime. However, scaling to millions of samples, may be infeasible with the current implementation of the mixed linear model approach, necessitating alternative methods designed for massive cohorts like SLEMM or ssGBLUP [38] .

Determining the rank of the Woodbury basis from the GRM, using any of the data-driven selection methods, can quickly become its own bottleneck. However, the effective rank (assuming a sufficiently large cohort) should be a relatively stable property of the breed/population and could be calculated once or estimated as the number of independent chromosome segments [8], and then reused in future GWAS. In practice, we found (on this simulated data) that an effective rank close to 90% of the eigenmass was within 1% of the exact mode, substantially lower than the theoretical 98/99% eigenmass [8].

“Warm-starting” the variance components with the HE estimator can reduce the number of REML iterations needed compared to the naïve equal partition for low heritability traits but has limited advantage at higher heritability. Given many relevant traits in livestock have low-to-moderate heritability [39], this leads to large savings on average. Traits with complex inheritance, i.e. with epistatic effects like Trait B, are poorly captured by the simple genetic+environment model and often fall back to the naïve estimator. Likewise, only damping the REML iteration steps when the underlying Hessian disagrees with the previous step expectation allows much more rapid convergence on all tested datasets, easing one of the GWAS major bottlenecks.

Detecting dominance effects at individual SNPs typically requires significantly larger cohorts to achieve statistical power compared to additive effects, and so dominance effects are becoming increasingly important to study as more genotype/phenotype data is collected. Despite being a simple recoding of genotypes, many GWAS tools (with a few exceptions like PLINK and REGENIE) primarily only support additive testing. As such, drastically simplifying and accelerating the dominance recoding in GCTAg will allow testing at larger scales, facilitating the discovery of trait-associated loci with potentially substantial dominance or recessive effects that impact agricultural populations.

**Conclusion**

GCTAg removes key computational bottlenecks, enabling exact full-GRM GWAS with six-figure sample sizes and millions of variants at scales previously infeasible with GCTA. An unbiased trace approximation further reduces computational requirements across all datasets, while related cohorts can exploit a reduced-rank Woodbury representation for additional gains. All memory-intensive steps (GRM, PCA, and REML) support out-of-core methods to further reduce peak memory requirements. GCTAg also supports native genotype recoding for non-additive QTL analyses. Combined, these improvements will enable efficient full-GRM GWAS to match the continuing growth of modern mapping cohorts.

**Methods**

*Baseline GCTA GWAS*

The GRM is built from genotypes $x_{ij}$ for a variant with allele frequency $p_j$ as

$$G = \frac{1}{m} ZZ^{\top} \text{ where } Z_{ij} = \frac{(x_{ij} - 2p_j)}{\sqrt{2p_j(1-p_j)}}$$

The principal components of the GRM are defined as

$$G = Q\Lambda Q^{\top}$$

where $Q$ contains the eigenvectors and $\Lambda$ is the diagonal eigenvalue matrix. We model the phenotype as

$$y = C\alpha + g + e \text{ with } g \sim N(0, \sigma_g^2 G) \text{ and } e \sim N(0, \sigma_e^2 I)$$

with the covariate matrix $C$, such that the phenotype covariance matrix is

$$V = \sigma_g^2 G + \sigma_e^2 I$$

where the variance components $\sigma_g^2$ (genetic) and $\sigma_e^2$ (environmental) are estimated during REML. The null-model projection matrix is constructed as

$$P_0 = V^{-1} - V^{-1}C(C^{\top}V^{-1}C)^{-1}C^{\top}V^{-1}$$

Finally, the effect size and standard error for each variant is calculated as

$$SE(\hat{\beta}_j) = (x_j^T P_0 x_j)^{-\frac{1}{2}} \text{ and } \hat{\beta}_j = SE(\hat{\beta}_j)^2 \times x_j^{\top} P_0 y$$

for the phenotype vector $y$. The Wald statistic is $\left(\hat{\beta}_j / \mathrm{SE}(\hat{\beta}_j)\right)^2$, tested against a $\chi^2$ distribution with one degree of freedom.

*Blocked MLMA operations*

We can calculate MLMA statistics over blocks of variants rather than per-variant. Fixed effects are pre-adjusted as

$$P_0 y = V^{-1} y_{\text{adj}} \text{ where } y_{\text{adj}} = y - X((X^{\top}V^{-1}X)^{-1}X^{\top}V^{-1}y)$$

For a block of variants $X_{\text{block}}$, the effect size and standard error are then

$$\hat{\beta}_{\text{block}} = \frac{X_{\text{block}}^{\top} V^{-1} y_{\text{adj}}}{\text{diag}\,(X_{\text{block}}^{\top} V^{-1} X_{\text{block}})} \text{ and } \mathrm{SE}(\hat{\beta}_{\text{block}}) = \text{diag}(X_{\text{block}}^{\top} V^{-1} X_{\text{block}})^{-1/2}$$

with the division taken elementwise, and the Wald statistic is the same as before.

*Randomised SVD methods*

We implemented two approaches to calculating eigenvectors with rSVD to construct the leading eigensystem $(Q_k, \Lambda_k)$, including a small oversampling buffer ($k_{\text{ext}} = k + p$) to ensure accurate eigenvalues. Using $q$ power iterations and a Gaussian sketch matrix $\Omega$, we form

$$G^q\Omega = \tilde{Q}R$$

as an approximate subspace of $G$ where $\tilde{Q}^\top\tilde{Q} = I_{k_{\text{ext}}}$. We use that $G$ is symmetric to simplify the more general power iteration form $(GG^\top)^q G$. Additional power iteration rounds sharpen the approximation of $G$ and reduce spectral lag. We then orthonormalise and recover approximate eigenpairs with a Rayleigh-Ritz step

$$\tilde{Q}^\top G\tilde{Q} = Z\tilde{\Lambda}Z^\top$$

to take the leading $k$ eigenpairs of $\left(\tilde{Q}Z, \tilde{\Lambda}\right)$, which is of rank $k_{\text{ext}}$. Alternatively, we can use the Nyström approximation

$$G\Omega(\Omega^\top G\Omega)^\dagger\Omega^\top G$$

where the dagger represents the Moore–Penrose pseudoinverse for positive semidefinite matrices. The latter is more robust to poorly conditioned matrices but typically has a larger spectral lag to exact methods.

*Hutch++ trace approximation*

At each REML round, we need the trace of $PA_i$. We can estimate this with the Hutch++ approximation using $s$ probes as

$$\text{tr}(PA_i) \approx tr(Q^\top PA_iQ) + \frac{1}{s}\sum_{j=1}^{s} r_j^\top PA_i r_j$$

where $Q = \text{orth}(PA_iS)$ for the random sketch matrix $S \in \{\pm 1\}^{N\times s}$ and $r_j = (I - QQ^\top)g_j$ where $g_j$ are stochastic probes drawn (independently of $S$) from the same Rademacher distribution. We never explicitly materialise $P$ or $V^{-1}$: $P$ is applied to vectors through the projection from introduced earlier, and $V^{-1}$ through triangular solves with the Cholesky factor of $V$.

*Reduced-rank Woodbury representation*

By representing the GRM with $G \approx Q_k\Lambda_kQ_k^\top$, where $k$ is the reduced-rank, we can use the Woodbury matrix identity to solve for applications of the phenotype covariance matrix inverse as

$$V^{-1}v = \sigma_{\text{eff}}^{-2}(v - Q_kD_cQ_k^\top v)$$

where $D_c = \text{diag}(c_i)$, $c_i = \frac{\sigma_g^2(\lambda_i - \lambda_{\text{tail}})}{\sigma_e^2 + \sigma_g^2\lambda_i}$, and $\sigma_{\text{eff}}^2 = \sigma_e^2 + \sigma_g^2\lambda_{\text{tail}}$ based on the truncated tail terms

$$\lambda_{\text{tail}} = \frac{\text{tr}(G) - \sum_{i\le k}\lambda_i}{N-k} \text{ and } \tau^2 = \frac{\text{tr}(G^2) - \sum_{i\le k}\lambda_i^2}{N-k} - \lambda_{\text{tail}}^2$$

We can adjust the $\log|V|$ as

$$(N-k)\log\sigma_{\text{eff}}^2 + \sum_{i\le k}\log\left(\sigma_e^2 + \sigma_g^2\lambda_i\right)$$

We also reformulate the variant-scan to use the reduced-representation, now solving

$$x^\top V^{-1}x = \sigma_{\text{eff}}^{-2}\left(\|x\|^2 - \left\|D_c^{1/2}Q_k^\top x\right\|^2\right)$$

*GRM-based rank selection*

We can calculate the reduced-rank for a given GRM using three separate methods based on the eigenvalue spectrum. In all cases, we calculate a fixed number of eigenvalues upfront and then test the following criterion for the chosen method. If the criterion is not satisfied, we calculate double the number eigenvalues and repeat.

For Marchenko–Pastur, we use the number of samples *N* and the number of variants *m*, finding the number of eigenvalues larger than the "edge" eigenvalue from random matrix theory.

$$k_{\mathrm{MP}} = |\{i: \lambda_i > \lambda_+\}| \text{ where } \lambda_+ = \left(1 + \sqrt{N/m}\right)^2$$

For the eigenmass approach, we find the number of eigenvalues needed to reach a target mass $\rho_{\mathrm{target}}$ chosen by the user.

$$k_{\mathrm{EIG}} = \min\left\{k: \frac{\sum_{i=1}^{k} \lambda_i}{\mathrm{tr}(G)} \geq \rho_{\mathrm{target}}\right\}$$

We can optimise the eigenvalue expansion round for this mode, noting that the remaining eigenvalues can each only contribute at most $\lambda_k$ mass. We can improve the lower bound of the rSVD expansion as

$$k_{\mathrm{next}} = \max\left(2 * k, k + \left\lceil \frac{\left(\rho_{target} - \rho_k\right) tr(G)}{\lambda_k} \right\rceil\right)$$

For the tail variance approach, we find the smallest number of total eigenvalues, such that the remaining estimated variance in the truncated eigenspectrum is smaller than $\xi_{\mathrm{target}}$ chosen by the user.

$$k_{\mathrm{VAR}} = \min\left\{k: \sqrt{\frac{\Delta(k)}{\mathrm{tr}(G^2)}} < \xi_{\mathrm{target}}\right\} \text{ where } \Delta(k) = \left[\mathrm{tr}(G^2) - \sum_{i=1}^{k} \lambda_i^2\right] - \frac{\left[\mathrm{tr}(G) - \sum_{i=1}^{k} \lambda_i\right]^2}{N - k}$$

*Adaptive REML damping and convergence*

The AI Hessian matrix ($H_{AI}$) is already computed during AI REML but contains information about the state of REML iterations that is not yet used. The magnitude of the Newton decrement ($\delta^2$) in the optimization is roughly captured as $\delta^2 = U^\top H_{AI}^{-1} U$ where $U$ is the REML score vector. We refine the convergence criterion from $\Delta \log(\mathcal{L}) < 1 \times 10^{-4}$, where $\mathcal{L}$ is the likelihood estimator, to $\delta^2 < \epsilon$ for a chosen $\epsilon$.

We also redefine the damping of AI REML steps, where updates to the variance components are damped by a factor $\zeta$ in GCTA as

$$\zeta = \begin{cases} 10^{-\frac{1}{2}} & \Delta \log(\mathcal{L}) > 1 \\ 1 & \text{otherwise} \end{cases}$$

We instead use

$$\zeta = \begin{cases} 1 & \text{previous step trustworthy} \\ \min\left\{1, (2\delta^2)^{-\frac{1}{2}}\right\} & \text{otherwise} \end{cases}$$

Where trustworthy is defined as the realised gain after the current step is close to the predicted gain from the previous step (i.e. convergence is locally smooth) as $\frac{\Delta \log(\mathcal{L})_{\mathrm{realised}}}{\Delta \log(\mathcal{L})_{\mathrm{predicted}}} \gtrsim 0.75$. In cases where the previous step was not trustworthy and convergence is not locally smooth; we damp relative to the remaining Newton decrement rather than a fixed constant.

*Out-of-core GRM operations*

Almost all calculations involving the GRM ($A_0$) can be divided into subblocks. For example, the quadratic term of the REML score function

$$R_0 = y^\top P A_0 P y$$

can be accumulated over $B$ blocks as

$$R_0 = \sum_{b=1}^{B} (Py)_b^\top A_{0,bb} (Py)_b \; + \; 2 \sum_{b=1}^{B} \sum_{j>b} (Py)_b^\top A_{0,bj} (Py)_j$$

with contributions from the diagonals and upper-triangle off-diagonals counted once and twice respectively (as the GRM is symmetric).

*Genotype and phenotype simulation*

We used AlphaSimR v2.1.0 [40] to simulate a dairy cattle population, with the random seed "1". We generated founder haplotypes using the MaCS-based coalescent simulator for 1,000 non-inbred diploid individuals with the recommended bovine parameters. Each of the 29 simulated autosomes retained 600,000 segregating sites, resulting in a total of 17.4 million variants. We simulated two quantitative traits, based on the genetic architecture of two milk traits with moderate and low heritability respectively: trait A ($h^2$=0.08, additive) and trait B ($h^2$=0.35, additive+dominance+epistasis). We also simulated a trait based on height (trait C: $h^2$=0.58, additive+dominance) that was not used during selection. We expanded the population over 23 generations (5 random matings, 15 selecting on true genetic values, and 3 selecting on estimated breeding values), preserving the target genetic architecture but allowing heritability to drift. Only animals younger than 10 years of age were retained in the population, with a minimum mating age of two years. For each mating cycle, only top-ranked bulls based on the selection index were used. We then exported the final set of 152,522 individuals and genotypes into PLINK2 bfile format [41].

We created the subsets using PLINK2, randomly shuffling samples and variants and then subsetting to the desired level, such that an $N$=37.5k dataset was a proper subset of an $N$=75k dataset.

*GCTAg implementation*

We forked GCTAg from commit a66e042 of https://github.com/JianYang-Lab/GCTA and manually integrated equivalent changes from a previously modified version of GCTA used for the 1000BG projects from https://git.wur.nl/vande018/gcta/-/tree/dosages. Our fork is GPL3 compliant and available at https://github.com/ASLeonard/GCTAg. We used commit 2acf070 (tagged "v1.1.1 Simmental") as the GCTAg version during benchmarking.

We also created a separate branch within GCTAg (https://github.com/ASLeonard/GCTAg/tree/YangLab ) to use the generalized build updates, minimal changes required to use SIMD (fixing segmentation faults due to unaligned memory), and the save/load REML state feature as applied to the "base" GCTA using commit fdfe25a (tagged "GCTA v1.95.1 Baseline"), which we refer to as "GCTA (rebuilt)", This allowed us to benchmark the code more fairly to avoid outdated compilers/libraries from influencing the benchmarking or profiling. We also downloaded the v1.95.1 GCTA binary for Linux and macOS-

arm64 from https://yanglab.westlake.edu.cn/software/gcta/#Download for an unmodified comparison, which we refer to as "GCTA (download)". However, this version links the BLAS libraries in the wrong order on Linux, mostly affecting the MLMA step (**Supplementary Note 4**).

We also provide an optimized c++ implementation ("BED2MaCH") to convert between PLINK2 binary files and MaCH format used in this manuscript (https://github.com/AnimalGenomicsETH/GCTAg_benchmark).

*Benchmarking*
We benchmarked both GCTAg and the minimally modified GCTA on an Apple laptop and on the ETH Zurich Euler cluster. We measured runtime, memory, and I/O metrics with the Z shell "time" builtin and GNU time respectively. We ran five replicates to minimize variation across non-exclusive node use. We profiled the code using flamegraph (v0.6.14 https://github.com/flamegraph-rs/flamegraph) on Laptop and Server, as well as further profiling on Laptop using Instruments (version 26.6 17F113). When testing CPU architecture, we also used nodes with AMD EPYC 7742 and 512 GB RAM, AMD EPYC 7H12 and 256 GB RAM, and AMD EPYC 7763 and 256 GB RAM. We recompiled GCTAg on each node type (Zen 2, 3, and 4) separately to ensure valid use of the native optimisations.

We provide the scripts used to profile the different steps of the GCTA GWAS, benchmark the comparisons between GCTA and GCTAg, and the plotting scripts at https://github.com/AnimalGenomicsETH/GCTAg_benchmark.

**Ethics approval and consent to participate**
Not applicable

**Consent for publication**
Not applicable

**Availability of data and materials**
Data sharing is not applicable to this article as no real datasets were generated during the current study.

**Competing interests**
The authors declare that they have no competing interests.

**Funding**
A doctoral fellowship as part of the MSCA Doctoral Network BullNet was funded by the Swiss State Secretariat for Education, Research and Innovation (SERI). This study was financially supported by the Arbeitsgemeinschaft Schweizerischer Rinderzüchter (ASR), Zollikofen, Switzerland, and the Federal Office for Agriculture (FOAG), Bern, Switzerland.

**Author contributions**

ASL, NKK, and HP conceived the project. ASL implemented the changes to GCTA. ASL and NW tested the code. QH generated the simulation data. ASL and HP drafted the manuscript. All authors read and approved the final manuscript.

**Acknowledgements**
We thank Dr. Wim Gorssen for additional testing of the GCTAg software.

*Supplementary Table 1. Number of iterations until REML convergence across three traits with different heritability. Variance components were initialised using either the default GCTA behaviour (equal partition of variance) or warm-starting from a HE regression. Each REML iteration could be damped under the default GCTA behaviour or use the "adaptive" convergence based on the remaining curvature in the model. REML was run using the* N=*18,750 subset, and final heritability estimates from all methods were within 0.01% of each other. * indicates the HE estimation failed and the fallback to equal variance was used.*

| | **trait A ($h^2$=0.11)** | **trait B ($h^2$=0.46)** | **trait C ($h^2$=0.59)** |
|---|---|---|---|
| Equal+Damped | 13 | 10 | 8 |
| HE-start+Damped | 5 | 10* | 8 |
| Equal+Adaptive | 8 | 7 | 4 |
| HE-start +Adaptive | 3 | 7* | 3 |

*Supplementary Table 2. GRM density calculated from public datasets taken from the Genome Variant Map (GVM) through the NGDC. Higher proportion of off-diagonal elements > 0.125 indicate higher relatedness and hence sparse assumptions break down. Metrics can vary depending on upstream allele frequency filtering but are consistent that domesticated species are generally too related to use sparse GRM approximations.*

| **Species** | **Group** | **Bio Category** | **Mean Abs Value** | **Prop > 0.05** | **Prop > 0.125** |
|---|---|---|---|---|---|
| Arabidopsis | Plant | Wild, predominantly selfing | 0.02196 | 0.04896 | 0.01941 |
| Maize | Plant | Domesticated, outcrossing | 0.01866 | 0.04573 | 0.008146 |
| Potato | Plant | Domesticated, clonal/outcrossing | 0.1363 | 0.7610 | 0.4055 |
| Soybean | Plant | Domesticated, predominantly selfing | 0.07901 | 0.6095 | 0.1509 |
| Tomato | Plant | Domesticated, predominantly selfing | 0.3674 | 0.9135 | 0.7663 |
| Rice | Plant | Domesticated, predominantly selfing | 0.1090 | 0.7973 | 0.3793 |
| Buffalo | Animal | Domesticated, outbreeding | 0.3934 | 0.9881 | 0.9789 |
| Chicken | Animal | Domesticated, outbreeding | 0.04628 | 0.3300 | 0.04819 |
| Duck | Animal | Domesticated, outbreeding | 0.04576 | 0.3139 | 0.05314 |
| Goat | Animal | Domesticated, outbreeding | 0.04074 | 0.3276 | 0.01486 |
| Pig | Animal | Domesticated, outbreeding | 0.1473 | 0.7547 | 0.4582 |
| Sheep | Animal | Domesticated, outbreeding | 0.08554 | 0.8293 | 0.08823 |
| Human_1kGP | Human | Wild population, outbreeding | 0.01722 | 0.04439 | 0.0003803 |

*Supplementary Table 3. Heritability estimation from exact and approximate methods, tested on* N=*150k,* m=*600k. Over-truncation tends to drive V(G) up as the reduced basis also typically inflates the REML scoring function.*

| | **Exact** | **Hutch++** | **Woodbury (k=5k)** | **Woodbury (k=10k)** | **Woodbury (k=20k)** |
|---|---|---|---|---|---|
| V(G) | 0.473 | 0.474 | 0.500 | 0.474 | 0.472 |
| V(e) | 4.095 | 4.095 | 4.076 | 4.095 | 4.096 |
| $h^2$ | 0.1035 | 0.1037 | 0.1093 | 0.1037 | 0.1033 |

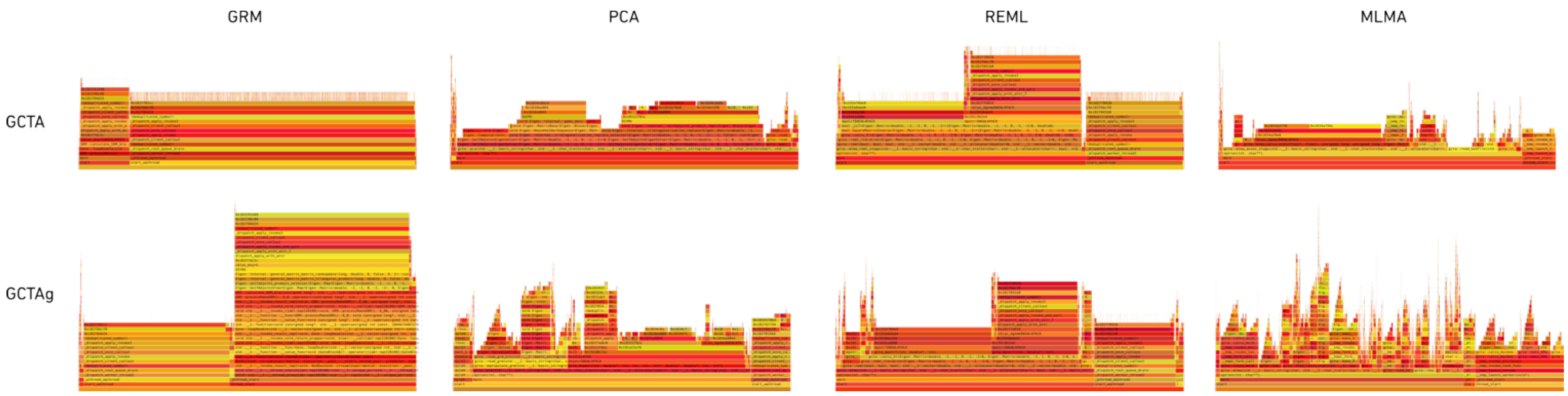

*Supplementary Figure 1. Flamegraphs from GCTA (rebuilt) and GCTAg on the profiling dataset for the four GWAS steps. We can identify potential bottlenecks in GCTA through the long calltime spent in e.g. PCA during the DSYMV operation or in MLMA in the SGEMV. Other steps like GRM and REML are mechanistically similar between GCTA and GCTAg.*

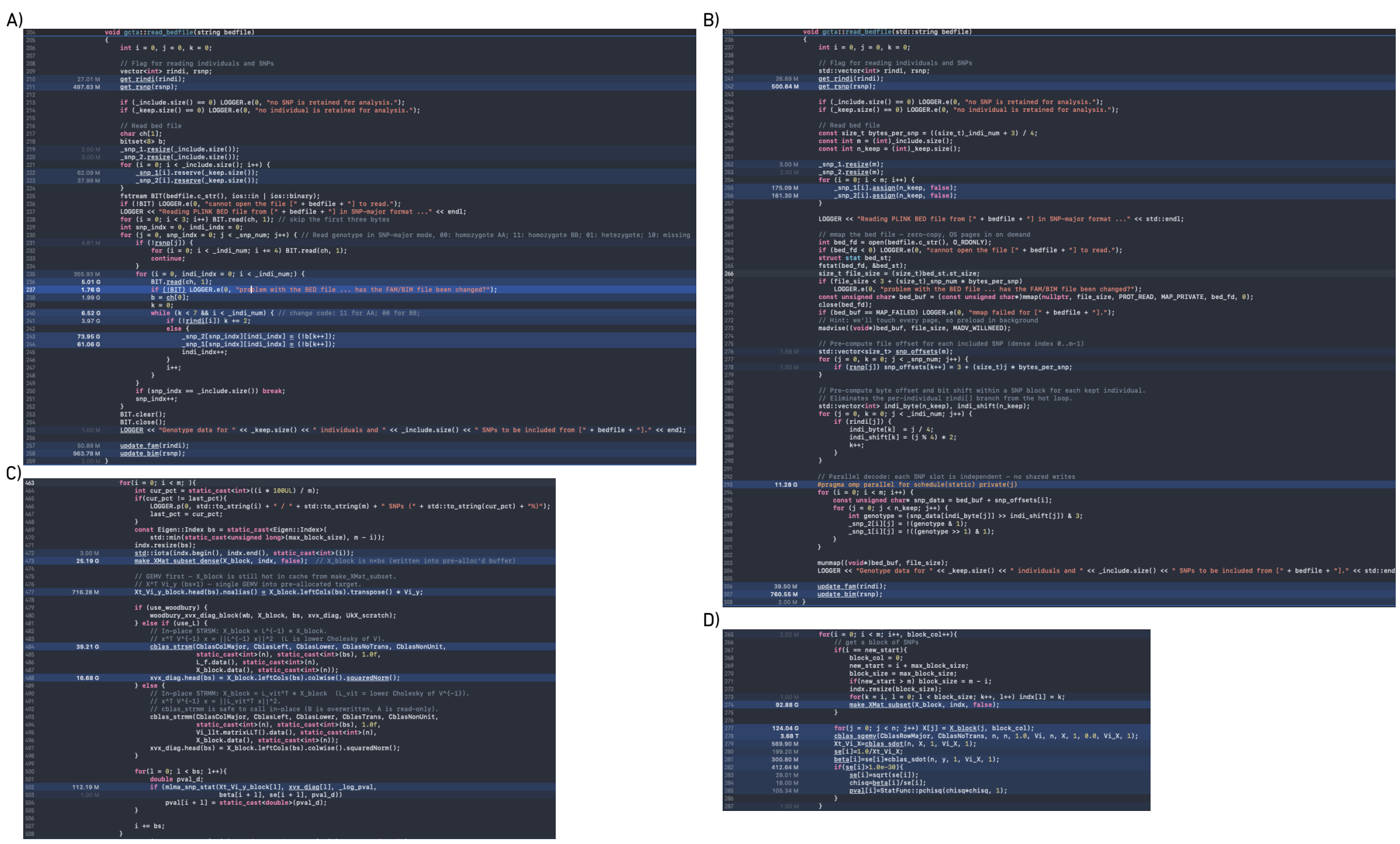

*Supplementary Figure 2. CPU cycle profiling. Byte-level I/O in GCTA (A) was slow, requiring 164.0 billion CPU cycles to read the BED file, compared to only 13.0 billion in GCTAg when using memory mapping and more efficient packed-genotype decoding. Likewise, the per-variant MLMA loop took 3.9 trillion CPU cycles (3.7 trillion in the SGEMV operation) in GCTA (C) compared to 84.7 billion (39.2 billion in the STRSM operation) in GCTAg (D).*

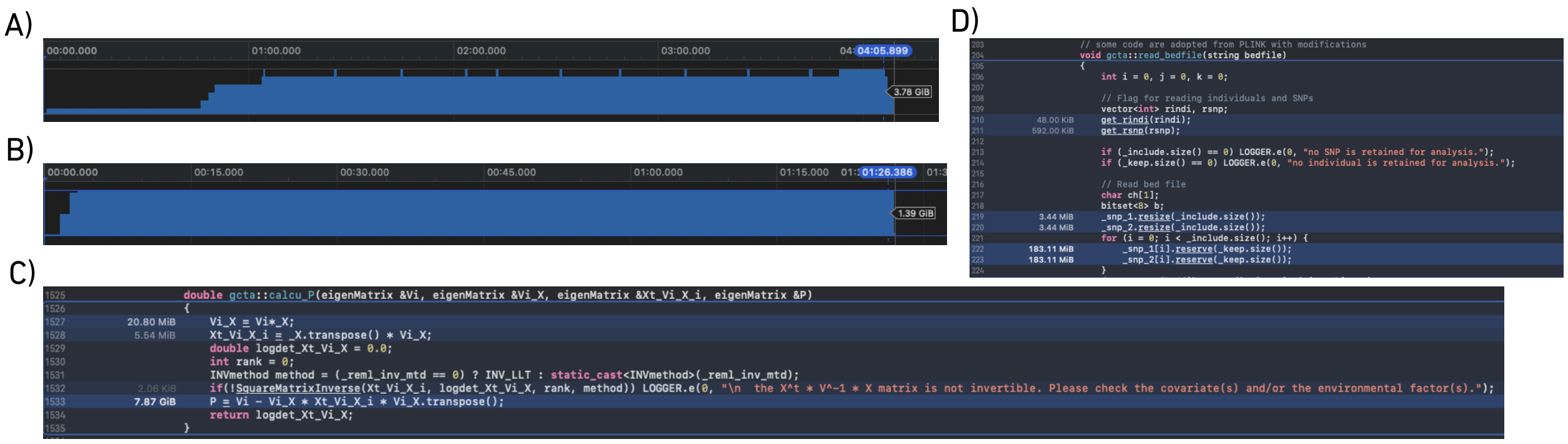

*Supplementary Figure 3. Memory allocation profiling of the REML step. Across the same dataset and program command, GCTA had a peak memory use of 3.8 GB with several spikes (A) compared to GCTAg having a smooth memory profile and peaking at 1.4 GB (~63% reduction). GCTA also allocated 35.1 GB of memory across the run, compared to only 13.7 GB for GCTAg, with large reallocations due to temporary matrices (C) or frequent reallocation of memory inside loops (D).*

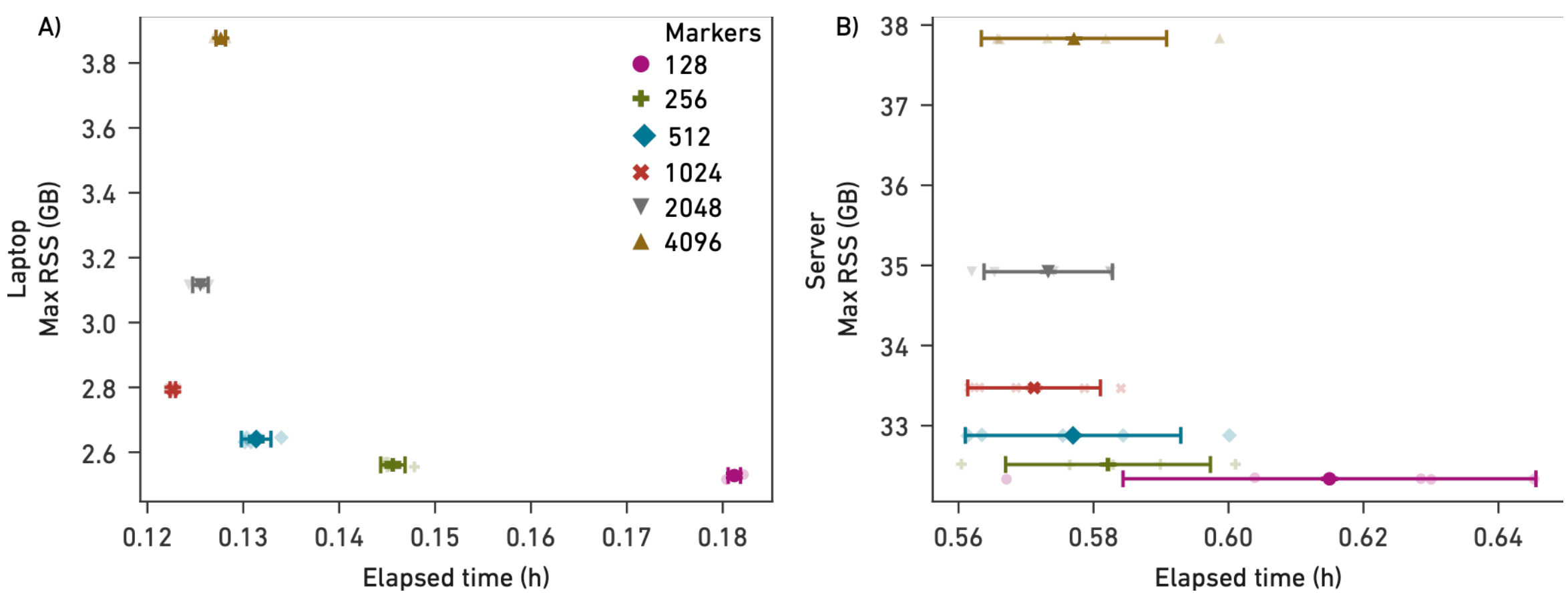

*Supplementary Figure 4. Increasing the number of variants per GRM accumulation batch from the hardcoded 128 value in GCTA improves throughput for a small increase in memory. Coincidentally, a batch size of 1024 variants is optimal when tested on Laptop (using N=18,750, m=600k) and Server (using N=75k, m=600k) despite their different BLAS libraries and CPU architectures.*

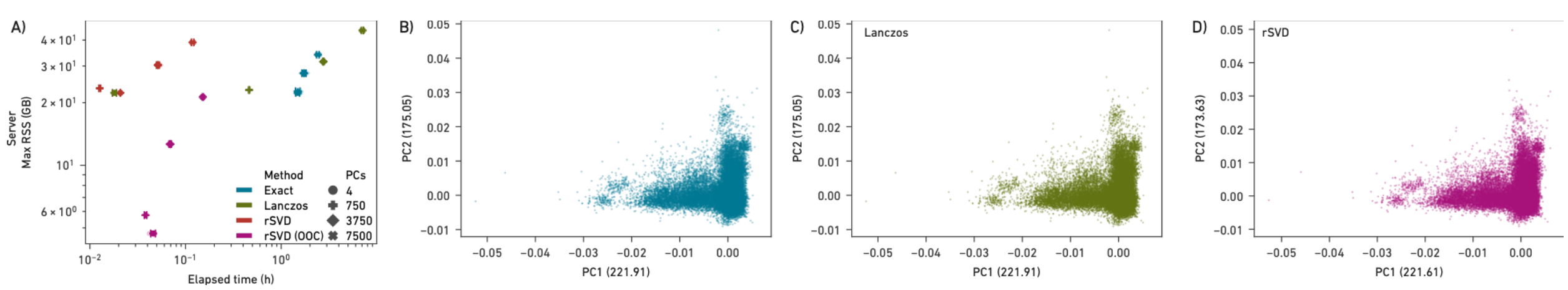

*Supplementary Figure 5. (A) The relative performance of eigensolver methods depends heavily on how many eigenvectors are needed, with e.g. Lanczos performing well when $k \ll n$ but worse than the exact solver once the spectral gap closes, run here for N=75k. The out-of-core (OOC) GRM budget was 2 Gb. All PCA methods are effectively indistinguishable for top PCs, with eigenvalues shown in parentheses, for (B) Exact (DSYEVR), (C) Lanczos, and (D) rSVD methods, although the rSVD has minor eigenvalue differences since it is not mathematically exact. The OOC rSVD method has identical output to the normal rSVD method.*

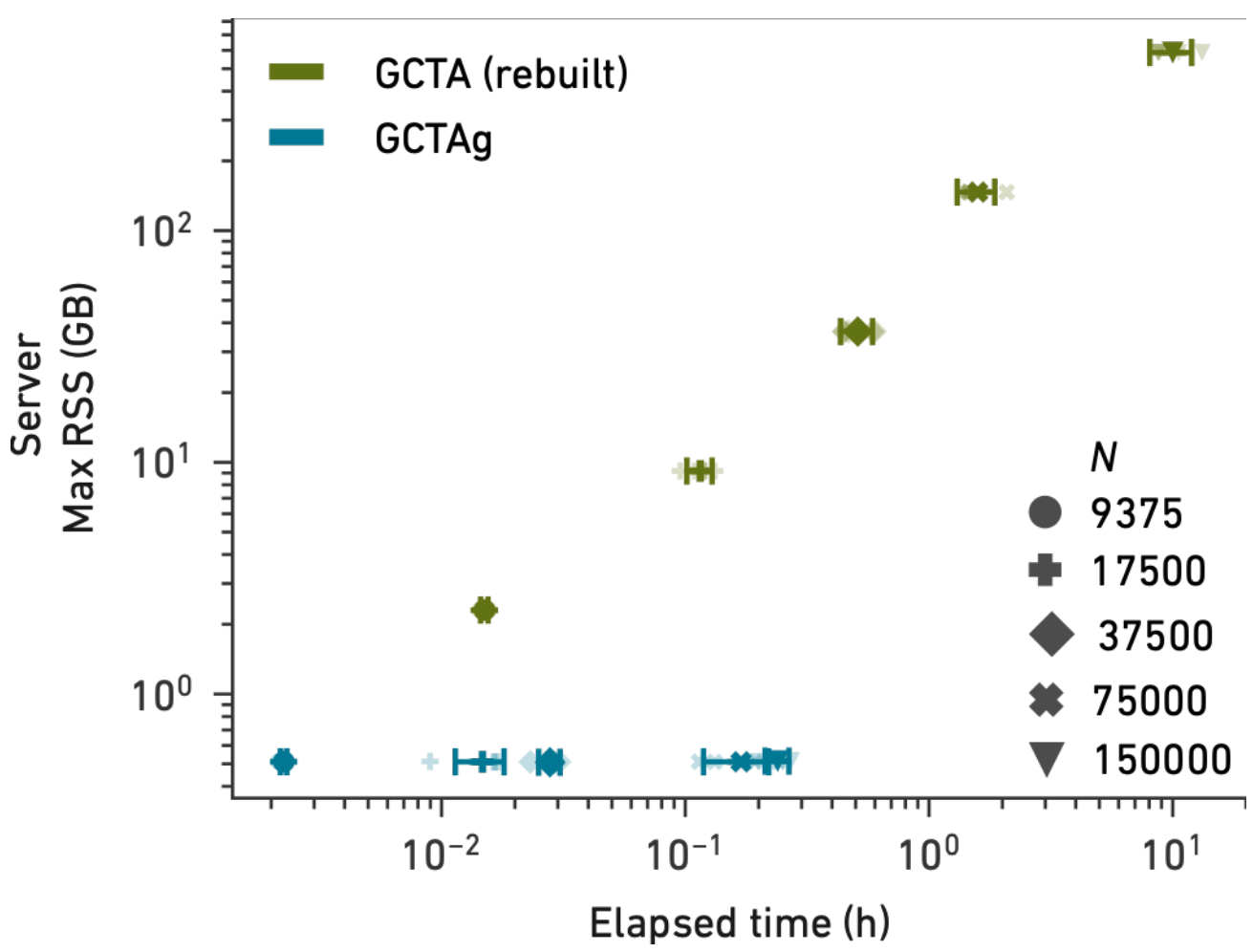

*Supplementary Figure 6. Merging GRMs from 29 autosomes scales quadratically with sample size for memory and time for GCTA but could be done in constant memory with GCTAg. Theoretically, GCTAg runtime should scale quadratically with sample size, but the short runtime was almost entirely I/O-bound and more susceptible to noise from other shared HPC jobs.*

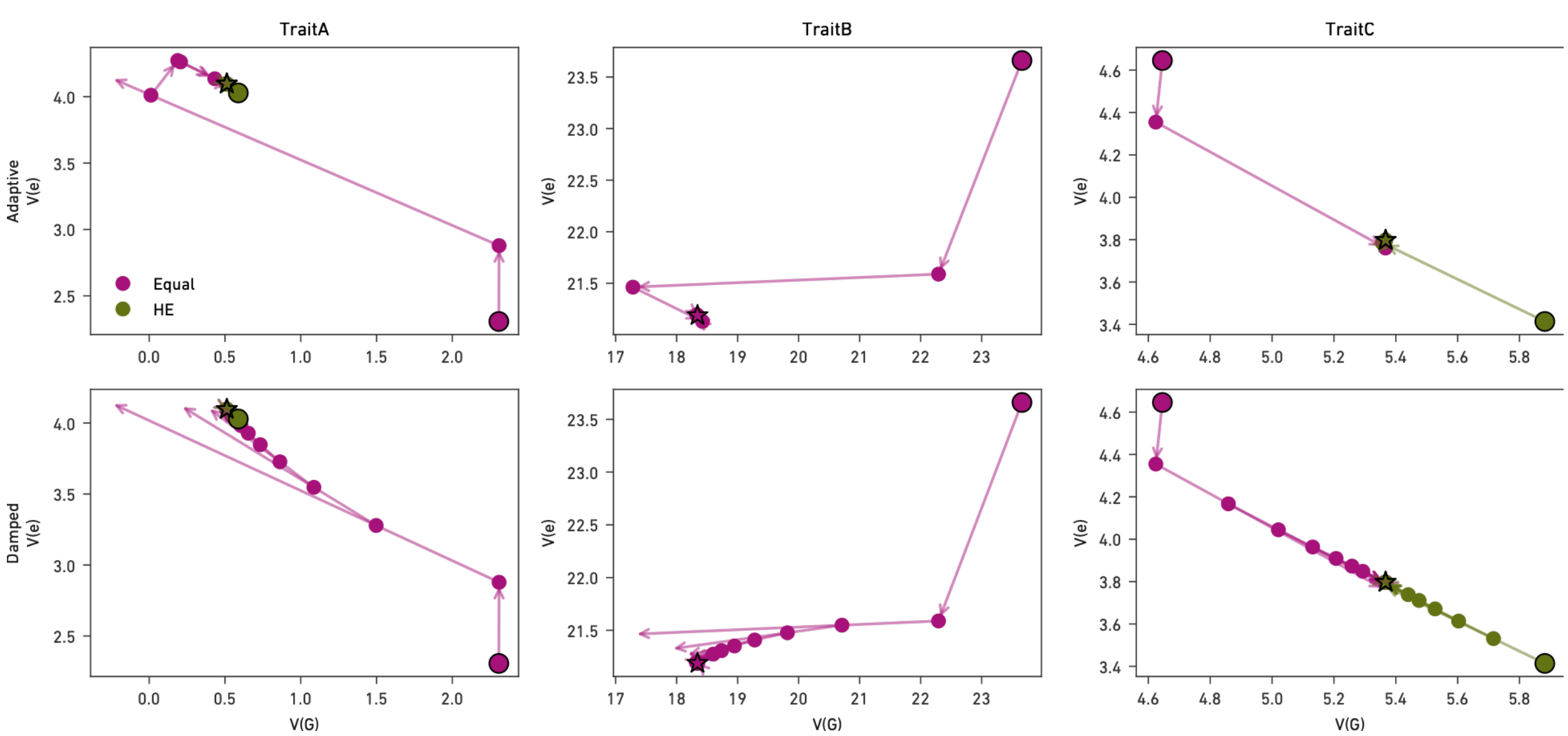

*Supplementary Figure 7. Across three simulated traits (see **Supplementary Table 1**), using either the HE-estimated or equal variance initialization (large circles) and damped or adaptive iterations all converged to effectively identical solutions (star). In general, the predicted steps are accurate (magnitude and direction shown by arrows) and damping steps by $\frac{1}{\sqrt{10}}$ based on an arbitrary absolute $\Delta\, log(\mathcal{L}) > 1$ condition is unnecessary. Only adaptively damping steps when the previous step was inaccurate, and otherwise taking the full step, converges much more quickly. Warm-starting Trait B failed and is not shown as it falls back to the equal partition initialisation. Note, the EM-REML first iteration for the equal start in the original GCTA implementation is never damped.*

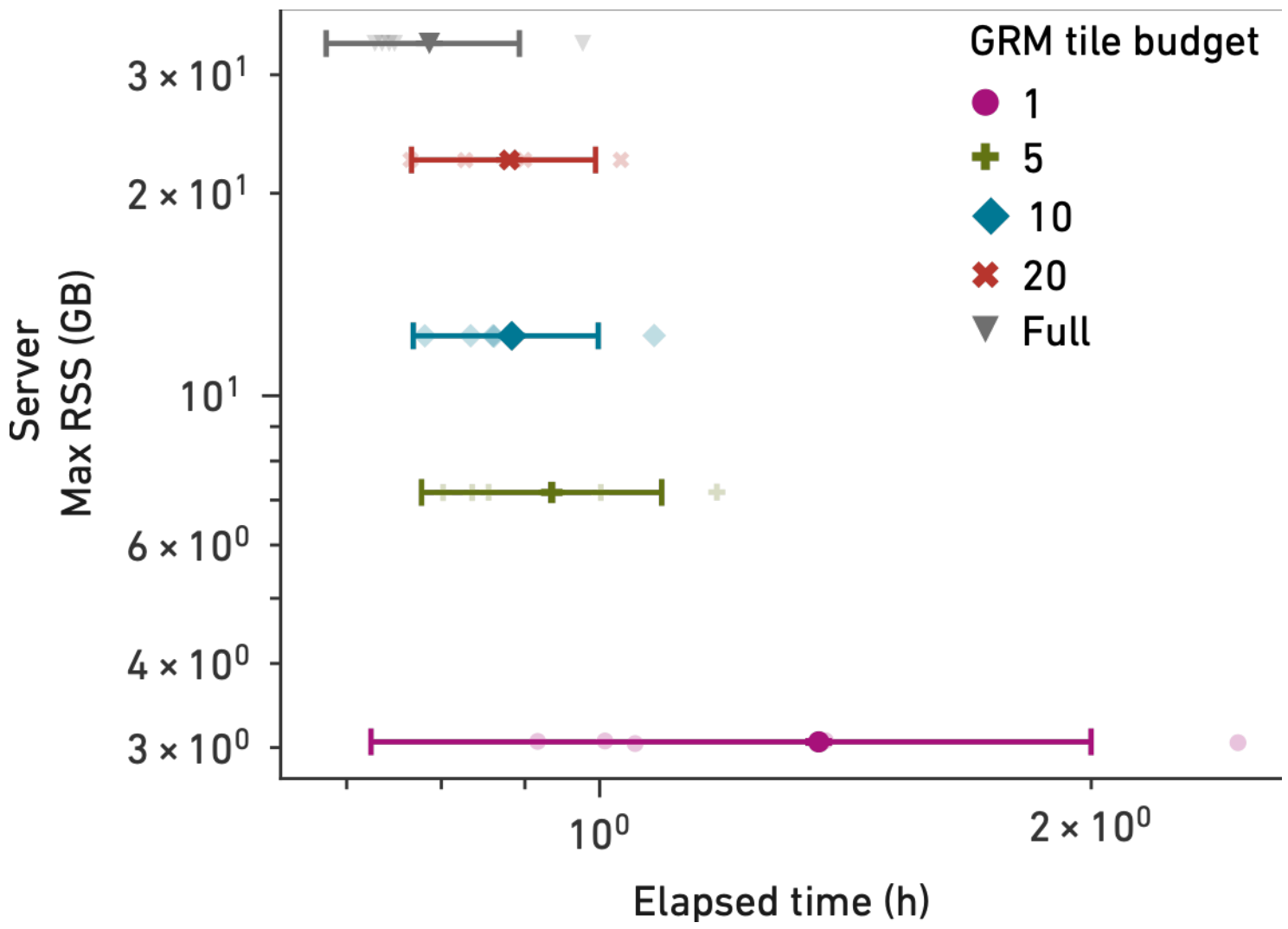

*Supplementary Figure 8. The GRM can be calculated with a tiled out-of-core approach, with multiple passes over the genotypes to calculate entries for different tiles of samples, run here for* N=75k, m=600k*. Using a small budget requires more I/O passes, which can have a small or large effect depending on I/O bandwidth relative to CPU bandwidth as well as I/O pressure.*

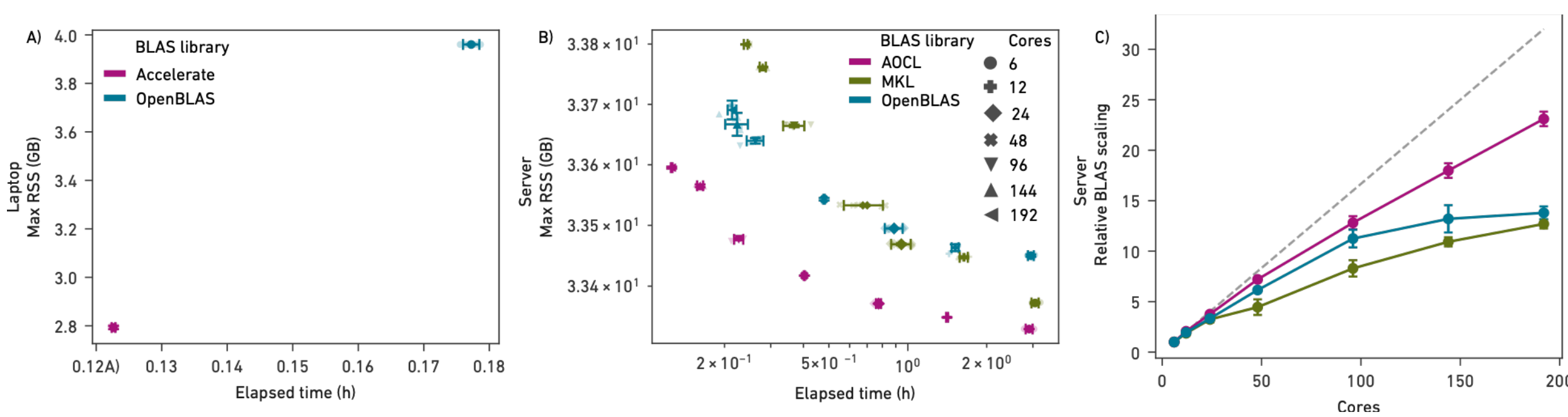

*Supplementary Figure 9. GRM building runtime on identical data can vary substantially depending on choice of BLAS library and core count. (A) For* N=18,750*,* m=600k *on Laptop, the macOS-native Accelerate was substantially better than OpenBLAS. Accelerate uses all cores (and the Apple silicon AMX co-processor), and so we only ran OpenBLAS using all 10 cores for a fair comparison. Most of the performance difference is due to OpenBLAS only using optimised CPU paths and not the AMX co-processor. (B) For* N=75k*,* m=600k *on Server, there is minimal difference in absolute runtime or memory across the BLAS libraries at low core count. However, the differences become substantial at high core counts when architecture-specific kernels can dominate. (C) The multicore scaling for each library (normalized by runtime at 6 cores for each library separately) highlights the AMD-tuned AOCL can leverage many cores on the AMD nodes more efficiently compared to Intel-tuned MKL and generic OpenBLAS which start to plateau at lower core counts. The dashed diagonal line indicates perfect scaling*

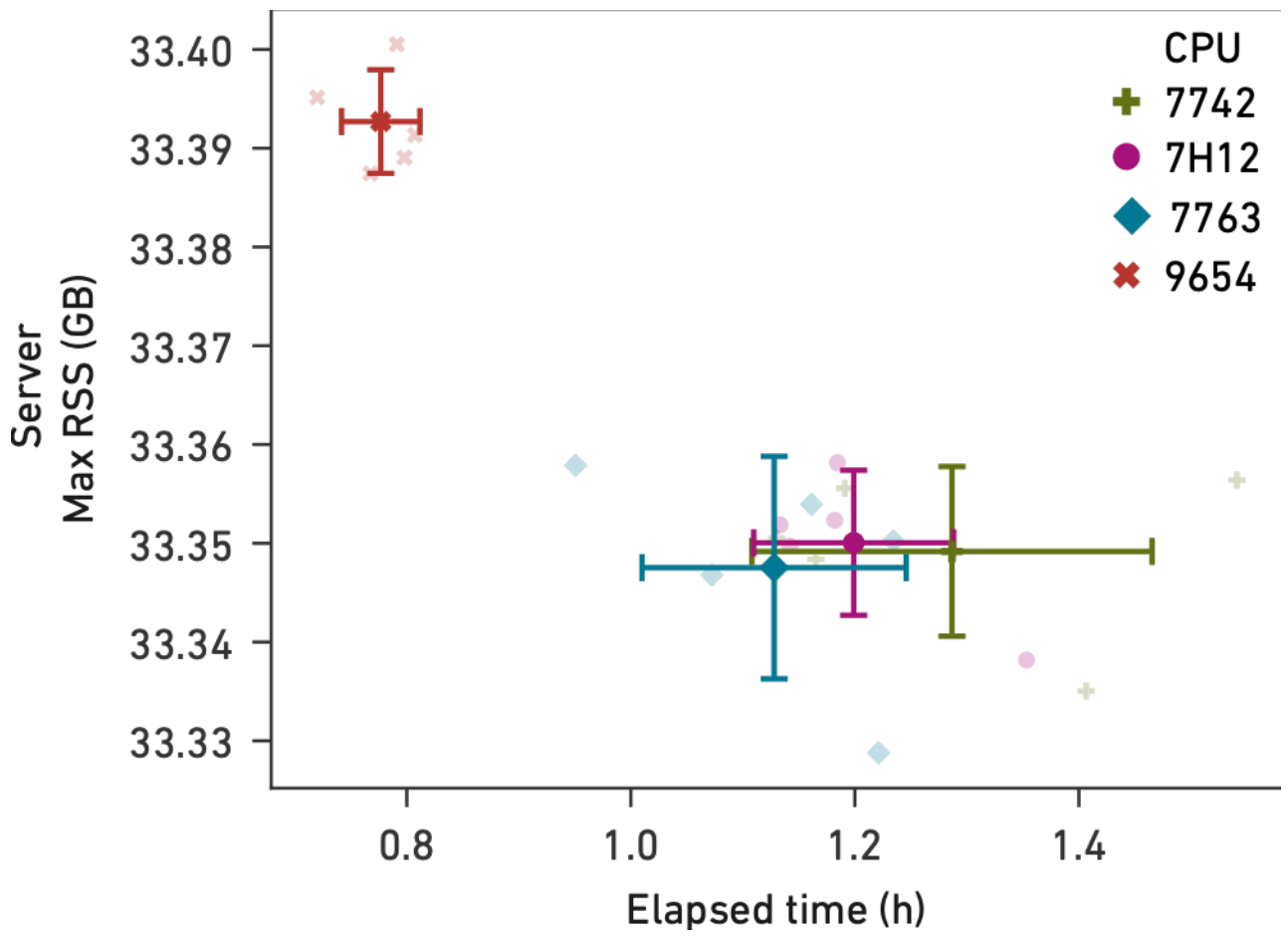

*Supplementary Figure 10. GRM building runtime on identical data (*N*=75k,* m*=600k) can vary substantially depending on CPU architecture. CPU architecture is especially important depending on SIMD availability. The 9654 CPU supports AVX-512, theoretically doubling FLOPs, while earlier generations can show slight improvements based on higher clock speed (7H12) or IPC/CCD (7763) over the baseline (7742).*

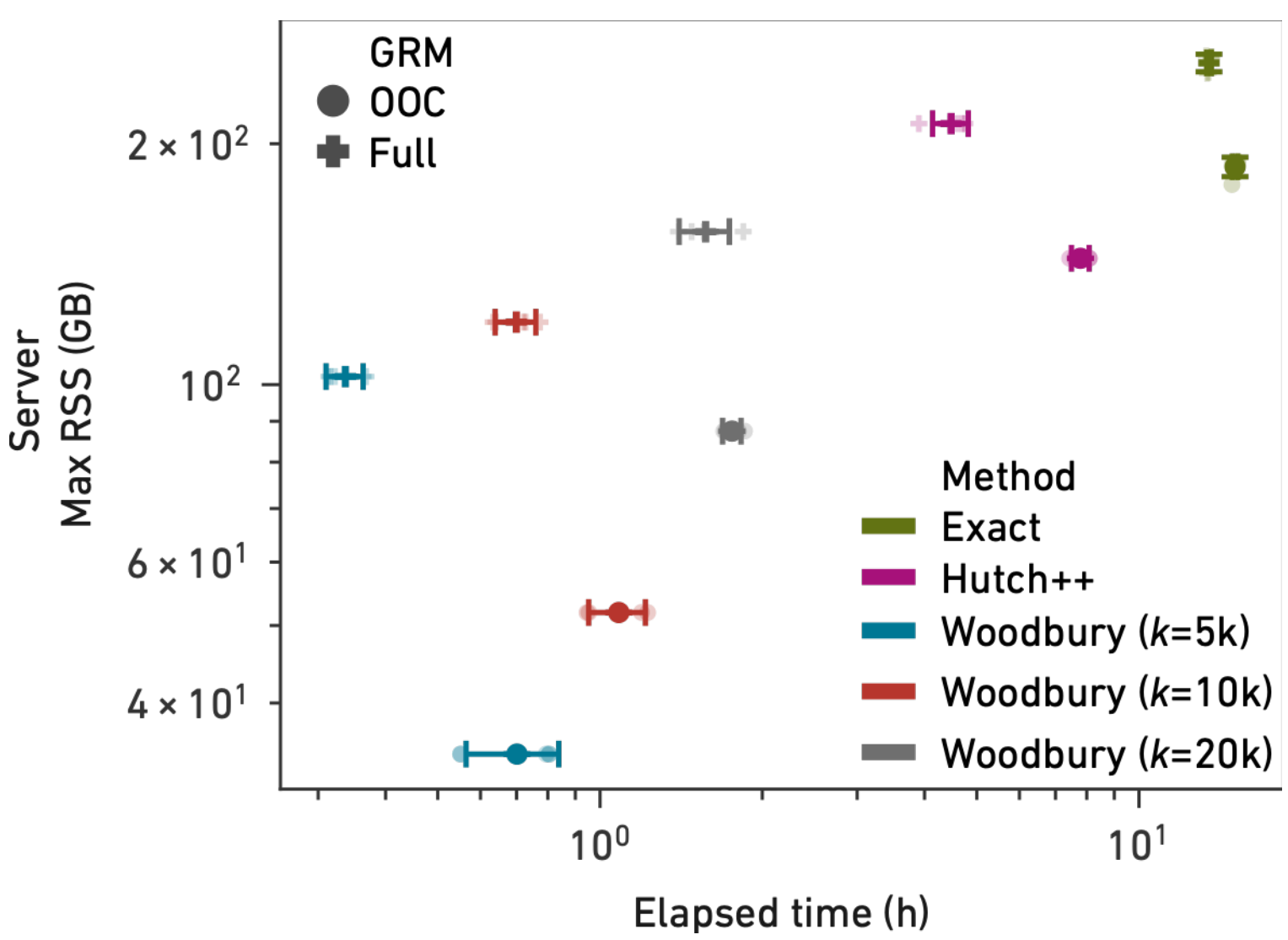

*Supplementary Figure 11. All REML modes (Exact, Hutch++, and Woodbury) can leverage out-of-core (OOC) GRM operations to calculate their relevant REML quantities, reducing a dominant quadratic cost to memory, shown here for* N*=150k,* m*=600k. Woodbury only needs to process the GRM once during the basis calculation, whereas Hutch++ and Exact must process the entire GRM each REML round and tend to incur a greater performance penalty. The "Full" GRM uses roughly 90 Gb at this sample size, whereas the OOC GRM budget was 10 Gb.*

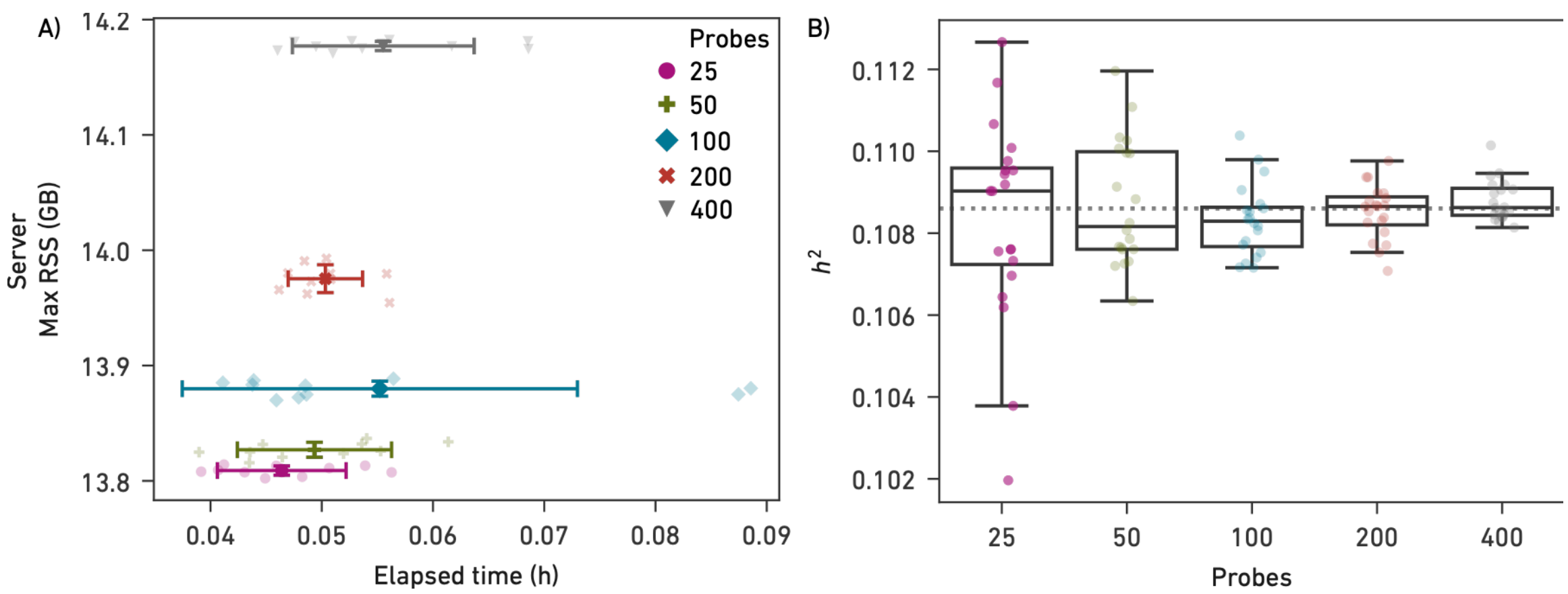

*Supplementary Figure 12. (A) Using more probes used during Hutch++ REML increases peak memory slightly but is a minor contributor to overall runtime, tested on (*N*=37.5k, *m*=600k). (B) The standard error of the heritability estimate (across separate runs) decreases with more probes, although the Hutch++ estimator is still unbiased and the mean is fairly stable. Note, this is not the standard error estimated for heritability within a single run, which is independent of the number of probes.*

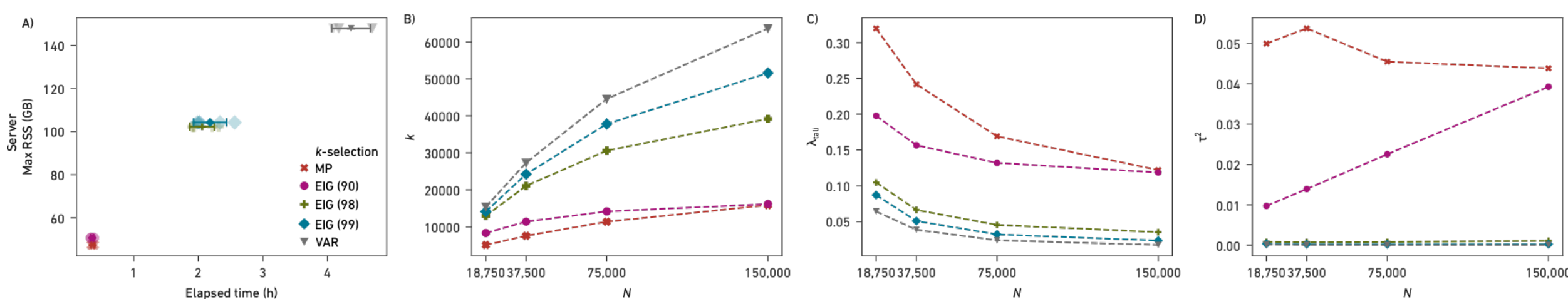

*Supplementary Figure 13. (A) The runtime and memory requirements of selecting the reduced rank size* k *from the data is dominated by the rSVD expansion rounds necessary to calculate sufficient eigenvalues, shown here for* N*=75k,* m*=600k using 48 cores. For example, both the EIG (90) and MP methods calculated approximately 20k eigenvalues before satisfying their convergence criteria, whereas both EIG (98) and EIG (99) ended up calculating nearly 54k eigenvalues. (B) The selected reduced rank generally grows sub-linearly, with more lenient thresholds like EIG (90) appearing nearly flat from* N*=75k to* N*=150k. The (C) $\lambda_{tail}$ and (D) $\tau^2$ terms from the truncated eigenspectrum can be used in a post hoc correction (see* ***Supplementary Note 4****). The terms becoming vanishingly small for conservative thresholds, indicating the reduced rank captures almost all non-degenerate information and the correction becomes negligible.*

## Supplementary Note 1

Symmetric problems, like the matrix multiplication in the GRM building where i.e. $G_{ij} = G_{ji}$ by design, can exploit symmetric matrix operations. In the case of the GRM, we can use DSYMM rather than DGEMM to only calculate the multiplication of one triangle rather than the whole square matrix. Although this is theoretically more efficient, not all BLAS routines are equally optimised and DGEMM in particular is thoroughly optimised. As such, DGEMM outperforms DSYMM on synthetic matrix data (**Table SN1**) despite the theoretical advantages of the symmetric version. However, reduced peak memory (and memory traffic) may be a net win depending on exact hardware configurations.

*Table SN1. Across different core counts and square matrix sizes, DGEMM routinely outperforms DSYMM despite the heavier memory traffic. BLAS operations are measures in runtime seconds and averaged over five repeated warm-start runs.*

| Cores | Size | DGEMM | DSYMM | DSYMM Overhead (%) |
|---|---|---|---|---|
| 8 | 20k | 37.9 | 38.5 | 1.8% |
| 8 | 40k | 298.8 | 303.8 | 1.7% |
| 32 | 30k | 41.8 | 43.6 | 4.4% |
| 32 | 40k | 71.4 | 72.7 | 1.9% |

Even though we use square matrices to store the e.g., GRM, we can designate operations as triangular (with Eigen as e.g., `triangularView<Eigen::Upper>()` or the `UPLO` field in BLAS/LAPACK). By only filling the e.g. upper triangle of the matrix and performing triangle-aware matrix operations, the other triangle of the matrix remains untouched. So, while the "virtual" memory allocation accounts for the full square matrix, the "resident" memory (the actual memory constraint to consider) only accounts for roughly the triangle. MacOS uses a "pagesize" (the operating systems memory management unit) of 16 Kb, while Linux normally uses 4 Kb. Linux can also use "transparent huge pages" (THP) of 2 Mb, which is typically much more effective for large memory allocations like the GRM matrix. However, using THPs (which varies by cluster/computer configuration) typically means all the virtual memory for the untouched triangle regions is still "faulted in" and becomes resident. In the case of the Euler HPC we tested on, THP is enabled by default, and we saw no memory savings after implementing the triangle-aware approach. By changing the memory allocator from glibc to jemalloc, allowing us to disable THP (MALLOC_CONF="thp:never"), we recovered the expected memory savings. While there is a negative cost to disabling THPs, since tracking many small pages is less efficient than fewer large pages, the memory savings are significant at large sample sizes.

## Supplementary Note 2

The Eigen::SelfAdjointEigenSolver (when EIGEN_USE_LAPACKE is true in the build) dispatches to DSYEV. MKL has a non-reference build of DSYEV, utilizing more BLAS-3 operations like DTRMM, whereas the other tested backends like OpenBLAS, libFLAME (from AOCL), and Accelerate appear to implement the reference netlib-style algorithm, bottlenecked in the BLAS-2 DSYMV operation (**Figure SN1**). On the same profiling dataset from **Supplementary Figure 1**, MKL, OpenBLAS, and Accelerate respectively took 78, 459 (5.9x slower), and 321 (4.1x slower) seconds.

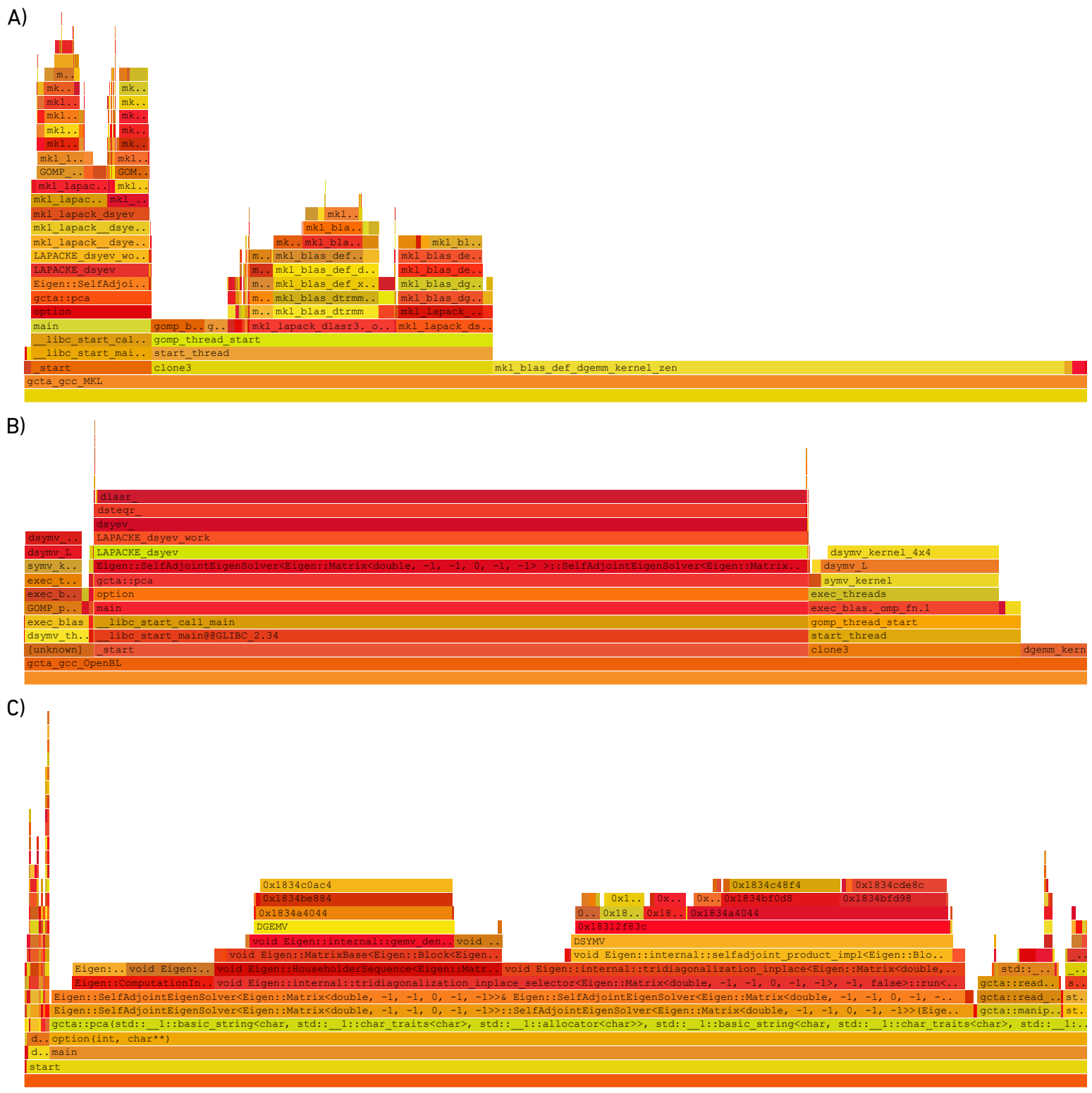

*Figure SN4. The Eigen SelfAdjointEigenSolver routing through DSYEV on Linux+MKL (A), Linux+OpenBLAS (B), and macOS+Accelerate(C). Although all calls start from DSYEV, they subsequently call different BLAS operations.*

Similarly, the D&C algorithm in DSYEVD requires a workspace quadratic in the number of samples. For the more common 32-bit "LP64" BLAS/LAPACK libraries, the workspace size should overflow the 32-bit integer if there are more than 32,766 samples, leading to a crash. The crash is observed in OpenBLAS, libFLAME, and Accelerate, but even the 32-bit LP64 MKL library appears to internally promote the workspace calculation to a 64-bit integer and can correctly compute the eigensystem. Both the DSYEVD and DYSEVR functions are also optimised/implemented differently in MKL and can be substantially faster than the AOCL or OpenBLAS versions (**Table SN2**).

*Table 2. Elapsed time for computing 100 (DSYEVR) or all (DSYEVD) PCs using 12 cores across two GRM sizes. Mean runtime is reported across five replicates with standard deviation shown in parentheses.*

| BLAS | GRM Size | DSYEVR | DSYEVD |
|---|---|---|---|
| AOCL | 9,375 | 69.5 (9.9) | 115.9 (24.8) |
| MKL | 9,375 | 27.0 (3.1) | 42.1 (6.8) |
| AOCL | 18,750 | 393.1 (33.4) | 516.3 (18.2) |
| MKL | 18,750 | 206.6 (5.2) | 187.2 (11.0) |

## Supplementary Note 3

After applying the reduced rank approach to the GRM, we can also examine the information remaining in the tail of the eigendecomposition $G = Q_k \Lambda_k Q_k^\top + \underbrace{\sum_{i=k+1}^{n} \lambda_i q_i q_i^\top}_{\text{tail},\ n-k\ \text{terms}}$. We can cheaply summarize two moments of the tail with $\lambda_{\text{tail}}$ and $\tau^2$ (see methods). During REML, we are solving for the root of the score function $U(\hat{\sigma}_g^2) = -\frac{1}{2}\text{tr}(PK) + \frac{1}{2}y^\top PKPy = 0$.
We can use a second-order Taylor expansion to correct the first term using the tail moments $\text{tr}(PK)_{\text{tail}} \approx (n-k)\left[\frac{\lambda_{\text{tail}}}{\hat{\sigma}_g^2\lambda_{\text{tail}}+\hat{\sigma}_e^2} - \hat{\sigma}_g^2\hat{\sigma}_e^2\frac{\tau^2}{(\hat{\sigma}_g^2\lambda_{\text{tail}}+\hat{\sigma}_e^2)^3}\right]$,
but the quadratic second term

$$y^\top PKPy|_{\text{tail}} = \sum_{i=k+1}^{n} \lambda_i (q_i^\top Py)^2$$

depends on the covariance structure between the tail eigenvalues and the squared alignment of the *unknown* tail eigenvectors. Estimating this covariance would require additional computation and would still contribute an inexact post hoc correction term. Applying only the correction to the first term in the REML scoring frequently led to rapid unbound divergence. As such, there is limited value in attempting to correct for the truncation, as increasing *k* directly is relatively cheap and proven to improve accuracy.

## Supplementary Note 4

The default Linux build (downloading v1.95.1 from https://yanglab.westlake.edu.cn/software/gcta/#Download) links GSL BLAS before MKL BLAS. While many linear algebra operations like DGEMM (used in GRM construction) still fall through to MKL, the SGEMV and SDOT calls (used in the MLMA variant-scan step) go through the scalar GSL implementation rather than MKL, leading to a catastrophic degradation of performance (**Figure SN2**). By setting `LD_PRELOAD` to force loading the MKL SGEMV and SDOT routines before GSL's versions, we can recover the performance. Alternatively, altering the linking order of libraries during compilation, such that MKL is given precedence over GSL also recovers the performance. The Apple version of the downloaded binary prioritises Accelerate over GSL, and so does not exhibit this issue.

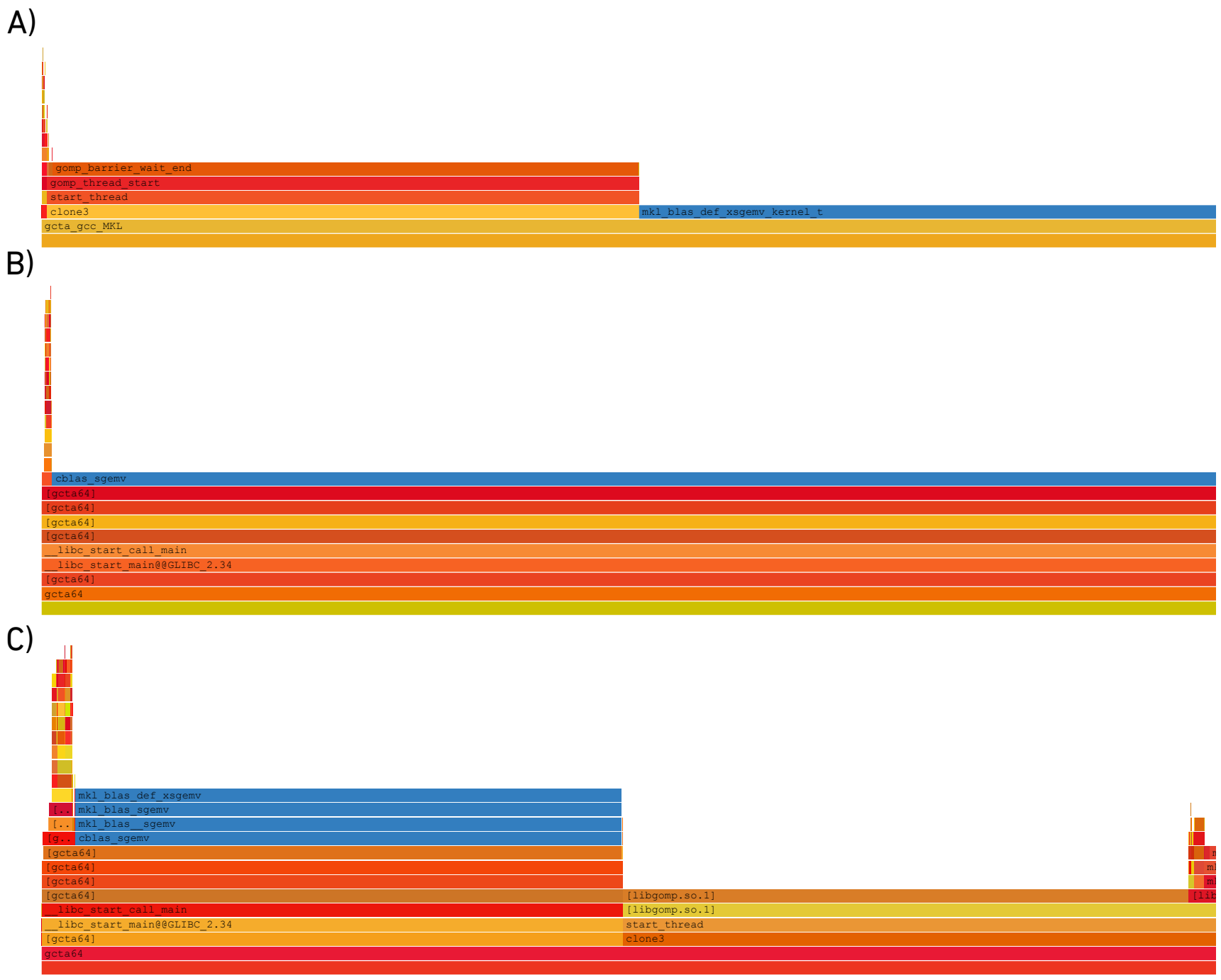

*Figure SN5. (A) The MLMA step in GCTA (rebuilt) primarily calls the SGEMV operation (marked in blue), correctly dispatched to MKL. (B) However, GCTA (download) on Linux links GSL before MKL and first resolves the SGEMV operation to the much slower/scalar GSL implementation. (C) Using LD_PRELOAD=<provided MKL> resolves the issue by prioritising the MKL version of SGEMV again. The macOS version correctly prioritises the Accelerate version on both the downloaded or self-compiled version and did not face this issue.*